\documentclass[10pt,american,prd,notitlepage,nofootinbib]{revtex4-1}
\usepackage[T1]{fontenc}
\usepackage[latin9]{inputenc}
\usepackage{bm}
\usepackage{amsmath}
\usepackage{amssymb}
\usepackage{graphicx}
\usepackage{esint}

\makeatletter
\@ifundefined{textcolor}{}
{%
 \definecolor{BLACK}{gray}{0}
 \definecolor{WHITE}{gray}{1}
 \definecolor{RED}{rgb}{1,0,0}
 \definecolor{GREEN}{rgb}{0,1,0}
 \definecolor{BLUE}{rgb}{0,0,1}
 \definecolor{CYAN}{cmyk}{1,0,0,0}
 \definecolor{MAGENTA}{cmyk}{0,1,0,0}
 \definecolor{YELLOW}{cmyk}{0,0,1,0}
 }

\def\Mpl{M_{\rm Pl}}

\usepackage{babel}

\usepackage{babel}

\newcommand{\be}{\begin{equation}}
  \newcommand{\ee}{\end{equation}}
\newcommand{\ba}{\begin{eqnarray}}
  \newcommand{\ea}{\end{eqnarray}}

\usepackage{babel}

\usepackage{babel}

\usepackage{babel}

\makeatother

\usepackage{babel}
\begin{document}

\title{Stability of the 3-form field during inflation}

\author{Antonio De Felice}

\affiliation{ThEP's CRL, NEP, The Institute for Fundamental Study, Naresuan University,
Phitsanulok 65000, Thailand}

\affiliation{Thailand Center of Excellence in Physics, Ministry of Education,
Bangkok 10400, Thailand}

\author{Khamphee Karwan}

\affiliation{ThEP's CRL, NEP, The Institute for Fundamental Study, Naresuan University,
Phitsanulok 65000, Thailand}

\affiliation{Thailand Center of Excellence in Physics, Ministry of Education,
Bangkok 10400, Thailand}

\author{Pitayuth Wongjun}

\affiliation{ThEP's CRL, NEP, The Institute for Fundamental Study, Naresuan University,
Phitsanulok 65000, Thailand}

\affiliation{Thailand Center of Excellence in Physics, Ministry of Education,
Bangkok 10400, Thailand}

\date{ }
\begin{abstract}
We consider the minimally coupled 3-form field which has been considered
as a candidate to realize inflation. We have studied the conditions
to avoid ghosts and Laplacian instabilities and found that some classes
of potentials, e.g.~the Mexican-hat one, will in general be unstable.
We then propose other classes of potentials which are instead free
from any instability, drive a long-enough slow-roll regime followed
by an oscillatory epoch, and as a consequence, can provide successful
inflation. Finally, we also provide stable potentials which lead to
a small enough propagation speed for the scalar perturbations, giving a possibility
for these models to produce non-Gaussianities.
\end{abstract}
\maketitle

\section{Introduction}

The inflation paradigm was introduced in 1980 as a way to solve different
issues, namely: the magnetic monopoles, the flatness, and the horizon
problems \cite{Guth:1980zm}; however, it can also account for the
observed temperature anisotropies in the Cosmic Microwave Background
(CMB) \cite{WMAP7} as well as the galaxy power spectrum \cite{LSS}.
In other words a sufficiently long stage of accelerated expansion
has been proposed as a way to solve all these problems at the same
time. In order to explain this period of accelerated expansion, some
new physics is introduced, and a scalar field \cite{Guth:1980zm,Linde:1983gd}
(or more than one \cite{Linde:1993cn,inflationmodels}) is commonly used. However,
the real mechanism for inflation is yet unknown, so it is interesting
to explore different possibilities which in general may lead to different
predictions for several inflationary observables (i.e.\ spectral
index, tensor-to-scalar ratio \cite{Smith:2005mm,Friedman:2006zt},
non-Gaussianities parameter \cite{Salopek,Gangui,Verde,KSpergel})
.

Since fundamental scalar fields have not been discovered yet in nature,
the idea of inflation might well be realized by other, higher-form
fields. For example, vector inflation (or one-form inflation) has
been intensively investigated \cite{Ford:1989me,Yokoyama:2008xw,Koivisto:2008xf,Golovnev:2008cf}.
Unfortunately, most of the vector-field models encounter instabilities
\cite{Himmetoglu:2008zp,Karciauskas:2008bc,Golovnev:2009ks,Himmetoglu:2009qi}.
More in general, the N-form field inflation has been also investigated
\cite{Germani:2009iq,Koivisto:2009sd} and one of the results is that
one-form and two-form fields are not stable, whereas the three-form
field can be stable \cite{Koivisto:2009sd}. Note also, that the four-form
field models correspond to the $f(R)$ theories \cite{Koivisto:2009sd,fofR}.

Recently, a form of inflation based on the evolution of a 3-form has
been studied \cite{Kobayashi:2009hj,Germani:2009gg,Koivisto:2009fb,Koivisto:2011rm}.
The origin of such a non-standard form for the inflaton may come from
a high-energy-scale theory, such as string-theory. Indeed it is interesting
to study such a model, as it may provide an alternative way to obtain
inflation. Since the essence of the 3-form is by construction different
from a single scalar field, we expect this difference to play some
role both at background and perturbation levels.

In fact, in this paper we will study the stability of a minimally
coupled 3-form during inflation with a general expression for the
potential. We will then find the conditions which avoid ghosts and
Laplacian instabilities (i.e.\ we require a positive kinetic term
and a non-negative speed of propagation for the independent linear
perturbation modes). Once these conditions are obtained, we reconsider
some models which have been recently introduced \cite{Koivisto:2009fb,Koivisto:2009ew},
and show that, if the potential is not carefully chosen, both ghosts
and Laplacian instabilities will occur.

Hence, we provide some classes of potentials, which, by construction,
are instead free from these instabilities, and, in this context, we
study their background evolution, in order to confirm that a slow-roll
period of inflation is then followed by a regime where the 3-form
oscillates, ending inflation. We also investigate about the possibility
of having stable evolutions and, at the same time, a small enough speed of
propagation for the scalar modes, opening the possibility of non-Gaussianities
signatures for these models. We will discuss the details of reheating,
and the bounds on the inflationary parameters (spectral index, tensor-to-scalar
ratio, and non-Gaussianities) in a future work.

The paper is organized as follows. In section \ref{sec:model}, we
introduce the Lagrangian of the model and write down the equations
of motion. Linear perturbation theory for this model is studied in
section \ref{sec:perturbation}, where we give the no-ghost conditions
and the squared speed of propagation for the scalar, vector and tensor
modes. We present some classes of potentials which make the model
free from ghosts and Laplacian instabilities in section \ref{sec:potential},
where we also show that a slow-roll period of inflation is followed
by an oscillatory regime which ends inflation. We write our conclusions
in section \ref{sec:fine}.

\section{The model and the background equations of motion\label{sec:model}}

Let us start with the following action
\begin{equation}
S=\int d^{4}x\sqrt{-g}\left[\frac{\Mpl^{2}}{2}\, R-\frac{1}{48}\, F_{\alpha\beta\gamma\delta}F^{\alpha\beta\gamma\delta}-V(A_{\alpha\beta\gamma}A^{\alpha\beta\gamma})\right],
\end{equation}
where $A_{\alpha\beta\gamma}$ is a 3-form, and $\bm{F}=\bm{d}\bm{A}$
is its Maxwell tensor \cite{MTW}, whose components can be written
as
\begin{equation}
F_{\mu\nu\rho\sigma}=\nabla_{\mu}A_{\nu\rho\sigma}-\nabla_{\sigma}A_{\mu\nu\rho}+\nabla_{\rho}A_{\sigma\mu\nu}-\nabla_{\nu}A_{\rho\sigma\mu}\,.
\end{equation}

\subsection{The background}

In this subsection, we review the background equations for three-form
field. All these equations and quantities were first derived
in \cite{Koivisto:2009sd}. Let us now consider a flat Friedmann-Lemaître-Robertson-Walker
(FLRW) manifold whose metric element can be written as
\begin{equation}
ds^{2}=-dt^{2}+a(t)^{2}d\bm{x}^{2}\,,
\end{equation}
and, on this background, considering Eq.~(\ref{eq:dual1}) in Appendix,
the background 3-form $A_{\alpha\beta\gamma}$ can be written as
\begin{equation}
A_{0ij}=0\,,\qquad A_{ijk}=a^{3}\epsilon_{ijk}\, X\,,
\end{equation}
where $\epsilon_{ijk}$ is the three-dimensional Levi-Civita symbol
(with $\epsilon_{123}=1$). Let us define the following quantities
\begin{equation}
V=V(y)\,,\quad V_{,y}\equiv\frac{dV(y)}{dy}\,,\quad V_{,yy}\equiv\frac{d^{2}V(y)}{dy^{2}}\,,\quad{\rm where}\quad y\equiv A_{\alpha\beta\gamma}A^{\alpha\beta\gamma}.
\end{equation}
On a FLRW background, we have $y=6X^{2}$. As a consequence of these
definitions, on FLRW, we have
\begin{eqnarray}
V(-X) & = & V(X)\,,\\
V_{,y} & = & \frac{dX}{dy}\, V_{,X}=\frac{V_{,X}}{12X}\,,\\
\dot{V} & = & 12X\dot{X}V_{,y}\\
V_{,yy} & = & \frac{1}{12X}\frac{d}{dX}\left(\frac{V_{,X}}{12X}\right)=\frac{XV_{,XX}-V_{,X}}{144X^{3}}\,,
\end{eqnarray}
so that we will restrict the form of the potential $V$ to even functions
of $X$. In this case the Friedmann equation can be written as
\begin{equation}
E_{1}\equiv3\Mpl^{2}H^{2}-\rho_{X}=0\,,\label{eq:fried}
\end{equation}
where
\begin{equation}
\rho_{X}=\frac{1}{2}\,\dot{X}^{2}+V+\frac{9}{2}\, H^{2}X^{2}+3HX\dot{X}=\frac{1}{2}\, Y^{2}+V\,,
\end{equation}
is the effective energy density of the 3-form, and we have defined
$Y\equiv\dot{X}+3HX$. The second Einstein equation reads as follows
\begin{equation}
E_{2}\equiv\Mpl^{2}(2\dot{H}+3H^{2})+p_{X}=0\,,
\end{equation}
where $p_{X}$ is the 3-form effective pressure defined as
\begin{equation}
p_{X}=-\left(\frac{1}{2}\,\dot{X}^{2}+V+3HX\dot{X}+\frac{9}{2}\, H^{2}X^{2}-12V_{,y}X^{2}\right)=2V_{,y}\, y-\rho_{X}\,.
\end{equation}
The equation of motion for the field gives
\begin{equation}
E_{X}\equiv\ddot{X}+3H\dot{X}+3X\dot{H}+12V_{,y}X=\dot{Y}+12X\, V_{,y}=0\,.\label{eq:eqX}
\end{equation}
The equations of motion are not all independent, due to Bianchi identities:
indeed we have
\begin{equation}
\dot{E}_{1}+3H(E_{1}-E_{2})+YE_{X}=0\,.
\end{equation}

One consequence of the equations of motion is
\begin{equation}
\Mpl^{2}\dot{H}=-V_{,y}y\,,\label{eq:supa}
\end{equation}
so that the universe will be super-accelerating when $V_{,y}<0$.

\section{Linear perturbation theory\label{sec:perturbation}}

\subsection{Scalar modes}

Let us consider now the metric for the scalar perturbations in the
following form \cite{cosmoper}
\begin{equation}
ds^{2}=-(1+2\alpha)dt^{2}+2\partial_{i}\psi\, dt\, dx^{i}+a^{2}(1+2\Phi)d\bm{x}^{2}\,,
\end{equation}
where we picked a spatial gauge so that the three-dimensional metric
is diagonal. As for the 3-form, by using once more Eq.~(\ref{eq:dual1})
given in Appendix, we can use a time gauge to fix the scalar perturbations%
\footnote{Here the field $\Phi$ in this gauge corresponds to the combination
$\Phi_{\mathrm{GI}}=\Phi-H\alpha_{0}/Y-HX(\partial^{2}\bar{\gamma})/Y$,
where, without fixing any gauge, $\alpha_{0}$ is defined, following
\cite{Koivisto:2009fb}, as $A_{ijk}=a^{3}\epsilon_{ijk}\,(X(t)+\alpha_{0})$,
and $\delta g_{ij}=a^{2}(2\Phi\delta_{ij}+2\partial_{i}\partial_{j}\bar{\gamma})$.
Then we can see that $\Phi_{\mathrm{GI}}$ is gauge invariant. In
other words, we have completely fixed the gauge freedom by setting
$\alpha_{0}=0=\bar{\gamma}$. This gauge-invariant field $\Phi_{\mathrm{GI}}$
is well defined as long as $Y=\dot{X}+3HX\neq0$. In particular, this
gauge is well defined in $X=0=y$, as long as its speed does not vanish,
that is $\dot{X}\neq0$ at $X=0$.%
} as \cite{Koivisto:2009fb}
\begin{equation}
A_{0ij}=a\epsilon_{ijk}\partial_{k}\beta(t,\bm{x})\,,\qquad A_{ijk}=a^{3}\epsilon_{ijk}\, X(t)\,.
\end{equation}

By expanding the action at second order in the fields we obtain
\begin{eqnarray}
S^{(2)} & = & \int dt\, d^{3}x\, a^{3}\left\{ \frac{6V_{,y}X^{2}}{a^{2}}(\partial\psi)^{2}-2\Mpl^{2}(H\alpha-\dot{\Phi})\frac{\partial^{2}\psi}{a^{2}}\right.\nonumber \\
 &  & {}+\frac{1}{2}\frac{(\partial^{2}\beta)^{2}}{a^{4}}+6V_{,y}\,\frac{(\partial\beta)^{2}}{a^{2}}+(Y\alpha+12V_{,y}X\psi+3Y\Phi)\,\frac{\partial^{2}\beta}{a^{2}}\nonumber \\
 &  & {}-\frac{1}{2}(6\Mpl^{2}H^{2}-Y^{2})\alpha^{2}+\left[6\Mpl^{2}H\dot{\Phi}-2\Mpl^{2}\frac{\partial^{2}\Phi}{a^{2}}+3(Y^{2}+12V_{,y}X^{2})\Phi\right]\alpha\nonumber \\
 &  & -\left.3\Mpl^{2}\dot{\Phi}^{2}+\Mpl^{2}\frac{(\partial\Phi)^{2}}{a^{2}}+\frac{9}{2}\left(Y^{2}-12V_{,y}X^{2}-144V_{,yy}X^{4}\right)\Phi^{2}\right\} .\label{eq:azione}
\end{eqnarray}
At a first look, this action has important differences with the general
action (for the perturbations) of scalar tensor theories \cite{gen2nd}.
First of all the presence of the terms $(\partial\psi)^{2}$ and $\Phi^{2}$
which, for a second-order general scalar-tensor theory, vanish after
using the equations of motion. Both these terms now vanish only when
the 3-form is absent. Furthermore the field $\beta$ is not dynamical
and it can be integrated out in Fourier space (together with $\alpha$
and $\psi$).

In order to remove these auxiliary fields it is convenient to work
in Fourier space: in this case, we can integrate out the fields $\alpha$,
$\psi$, and $\beta$, by using their own equations of motion. In
Fourier space, with $\Phi(t,\bm{x})=(2\pi)^{-3/2}\int d^{3}k\tilde{\Phi}_{\bm{k}}e^{i\bm{k}\cdot\bm{x}}$,
with reality condition $\tilde{\Phi}_{-\bm{k}}=\tilde{\Phi}_{\bm{k}}^{*}$,
the equations of motion for the constraints give
\begin{equation}
12V_{,y}X^{2}\psi+2\Mpl^{2}(H\alpha-\dot{\Phi})-12V_{,y}X\beta=0\,,
\end{equation}

\begin{equation}
(Y^{2}-6\Mpl^{2}H^{2})\alpha+\frac{2\Mpl^{2}Hk^{2}\psi}{a^{2}}+6\Mpl^{2}H\dot{\Phi}+2\Mpl^{2}\frac{k^{2}\Phi}{a^{2}}+3(Y^{2}+12V_{,y}X^{2})\Phi-\frac{Yk^{2}\beta}{a^{2}}=0\,,
\end{equation}
and
\begin{equation}
\frac{k^{2}}{a^{2}}\,\beta+12V_{,y}\beta-Y\alpha-12V_{,y}X\psi-3Y\Phi=0\,,
\end{equation}
where we omitted the tilde of the Fourier modes for simplicity. This
last equation can be solved for $\beta$ as
\begin{equation}
\beta=\frac{a^{2}(Y\alpha+12V_{,y}X\psi+3Y\Phi)}{k^{2}+12V_{,y}a^{2}}\,,
\end{equation}
so that we also have
\begin{equation}
\psi=\left(\frac{\Mpl^{2}}{6V_{,y}X^{2}}+\frac{2a^{2}\Mpl^{2}}{X^{2}k^{2}}\right)\dot{\Phi}+\frac{3a^{2}Y}{Xk^{2}}\,\Phi-\left(\frac{\Mpl^{2}H}{6V_{,y}X^{2}}+\frac{a^{2}(2\Mpl^{2}H-XY)}{k^{2}X^{2}}\right)\alpha\,,
\end{equation}
and finally
\begin{eqnarray}
\alpha & = & {\frac{\Mpl^{4}H{k}^{2}+6\,\Mpl^{2}V_{{,y}}a^{2}(3\, H{X}^{2}+2\,\Mpl^{2}H-XY)}{\Mpl^{2}H\,[\Mpl^{2}{k}^{2}H+6\, V_{{,y}}a^{2}(3\, H{X}^{2}+2\,\Mpl^{2}H-2\, XY)]}}\,\dot{\Phi}\nonumber \\
 &  & {}+{\frac{6\, V_{{y}}\Mpl^{2}{k}^{2}{X}^{2}+18\, V_{{,y}}a^{2}X\,(6\,{X}^{3}V_{{,y}}+\Mpl^{2}HY)}{\Mpl^{2}H\,[\Mpl^{2}{k}^{2}H+6\, V_{{,y}}a^{2}\,(3\, H{X}^{2}+2\,\Mpl^{2}H-2\, XY)]}}\,\Phi.
\end{eqnarray}
By substituting these expressions into the action written in Fourier
space, we find
\begin{equation}
\tilde{S}=\int dtd^{3}k\, Q(t,k^{2})\left[\dot{\Phi}_{\bm{k}}\dot{\Phi}_{-\bm{k}}-c_{X}^{2}(t,k^{2})\,\frac{k^{2}}{a^{2}}\,\Phi_{\bm{k}}\Phi_{-\bm{k}}\right].\label{action-s}
\end{equation}
If $Q>0$, we can define a canonical field $\Psi$ as
\begin{equation}
\Phi=\sqrt{\frac{a}{2Q}}\,\Psi\,,
\end{equation}
whose Lagrangian, in conformal time $\eta$, reads
\begin{equation}
\tilde{S}=\int d\eta d^{3}k\left[\frac{1}{2}\Psi_{\bm{k}}'\Psi_{-\bm{k}}'-\frac{1}{2}c_{X}^{2}k^{2}\Psi_{\bm{k}}\Psi_{-\bm{k}}-\frac{1}{2}\, m_{\Psi}^{2}\Psi_{\bm{k}}\Psi_{-\bm{k}}\right],
\end{equation}
where $m_{\Psi}^{2}$ is
\begin{equation}
m_{\Psi}^{2}=\frac{1}{2}\left(\frac{a''}{a}-\frac{{a'}^{2}}{a^{2}}-\frac{Q''}{Q}+\frac{{Q'}^{2}}{Q^{2}}\right)-\frac{1}{4}\left(\frac{a'}{a}-\frac{Q'}{Q}\right)^{2}\,,
\end{equation}
and a prime denotes differentiation with respect to conformal time.

\subsubsection{No-ghost conditions}

The no-ghost condition is then found to be
\begin{equation}
Q>0\,,\label{eq:noghost}
\end{equation}
where
\begin{equation}
Q={\frac{6{a}^{5}\Mpl^{2}V_{{,y}}{Y}^{2}}{\Mpl^{2}{k}^{2}H^{2}+6\, V_{{,y}}{a}^{2}\,(3\, H^{2}{X}^{2}+2\,\Mpl^{2}H^{2}-2\, XYH)}}\,.\label{eq:no-ghostQ}
\end{equation}
It should be noticed that condition (\ref{eq:noghost}) should hold
at all times during inflation, whether or not the trajectory is in
a slow-roll regime. On using the equations of motion (by replacing
$\Mpl^{2}H^{2}$ with the Friedmann equation and then $3HX=Y-\dot{X}$)
we find that
\begin{equation}
3\, H^{2}{X}^{2}+2\,\Mpl^{2}H^{2}-2\, XYH=\frac{1}{3}\,\dot{X}^{2}+\frac{2}{3}\, V\,,
\end{equation}
so that
\begin{equation}
Q={\frac{6{a}^{5}\Mpl^{2}V_{{,y}}{Y}^{2}}{\Mpl^{2}{k}^{2}H^{2}+2\, V_{{,y}}{a}^{2}\,(\dot{X}^{2}+2V)}}\,.
\end{equation}

This quantity must be positive for all $k$'s. For high $k$, we find
the condition $V_{,y}>0$. This condition must be satisfied along
the trajectory of motion. In some cases, it may be possible that for
some (positive) values of $y$, $V_{,y}$ is negative, but such values
of $y$ are never reached: in this case the model can still be viable.
We note here that the condition $V_{,y}>0$, on using Eq.~(\ref{eq:supa}),
forbids the dynamics to be super-accelerating. It is worth to note
that the condition $V_{,y}>0$ was also found in \cite{Koivisto:2009fb},
by demanding the background condition $p_{X}+\rho_{X}>0$. However,
we argue that for general theories the positivity of the sum of the
effective pressure and density does not necessarily imply the no-ghost
condition (\ref{eq:noghost}) (see also \cite{gen2nd}). In other
words condition (\ref{eq:noghost}) is a condition for the instability
of the perturbation modes, and not a condition on the background dynamics.

For low $k$'s we find another requirement, that is $\dot{X}^{2}+2V>0$.
Once more, this condition must be satisfied along the trajectory of
motion. The bottom line is that the two conditions $V\geq0$, $V_{,y}>0$
are sufficient conditions for not having ghosts. If these conditions
are not satisfied for all (positive) $y$'s, one should check that,
at least for the values of $y$ along the trajectory of motion for
the model, the above mentioned conditions still hold.

\subsubsection{Speed of propagation}

The speed of propagation is found as the large-$k$ limit of $c_{X}^{2}(t,k^{2})$
of Eq.~(\ref{action-s}). One can show that the speed of propagation,
on using the background equations of motion, is given as
\begin{equation}
c_{X}^{2}=\lim_{k\to\infty}c_{X}^{2}(t,k^{2})=1+\frac{2V_{,yy}y}{V_{,y}}=\frac{X\, V_{,XX}}{V_{,X}}\,.\label{cx2}
\end{equation}
The speed of propagation found here corresponds to the one found by
Koivisto and Nunes \cite{Koivisto:2009fb}. In general, only the simple
quadratic potential $V\propto y$, implies a propagation with speed
of light for all dynamics. Since $y\geq0$, then a sufficient condition
to avoid also Laplacian instabilities (besides the ghosts, $V_{,y}>0$)
is $V_{,yy}\geq0$.

\subsection{Vector modes}

Let us define the metric perturbation for the vector modes as
\begin{equation}
\delta g_{0i}=a\, G_{i}\,,\qquad{\rm and}\qquad\delta g_{ij}=a^{2}(C_{i,j}+C_{j,i})\,,
\end{equation}
where $G_{i,i}=0=C_{i,i}$. We will also choose a gauge for which
the 3-form has no vector perturbations (uniform field vector-gauge).
This choice completely fixes the gauge degrees of freedom. In this
case, one can show that the action for the vector modes becomes
\begin{equation}
S=\int dtd^{3}x\left[6a^{5}V_{,y}X^{2}\dot{C}_{i}\dot{C}_{i}+12a^{4}V_{,y}X^{2}\dot{C}_{i}Z_{i}+\frac{1}{4}\Mpl^{2}a\,(\partial_{j}Z_{i})(\partial_{j}Z_{i})+6a^{3}V_{,y}X^{2}Z_{i}Z_{i}\right],
\end{equation}
where we introduced the field $Z_{i}=G_{i}-a\,\dot{C}_{i}$. By introducing
Fourier modes, it is possible to integrate out the field $Z_{i}$
as
\begin{equation}
\tilde{Z}_{i}(t,\bm{k})=-\frac{24a^{3}V_{,y}X^{2}\dot{\tilde{C}}_{i}(t,\bm{k})}{\Mpl^{2}k^{2}+24a^{2}V_{,y}X^{2}}\,,
\end{equation}
so that the action for the vector modes becomes
\begin{equation}
S=\int dtd^{3}k\, Q_{V}(t,k^{2})\left[\dot{\tilde{C}}_{i}(t,\bm{k})\dot{\tilde{C}}_{i}(t,-\bm{k})\right],
\end{equation}
so that it is clear that the vector modes do not propagate.

\subsubsection{No-ghost condition}

The no-ghost condition for the vector modes corresponds to $Q_{V}>0$,
that is
\begin{equation}
Q_{V}=\frac{6k^{2}a^{5}\Mpl^{2}V_{,y}X^{2}}{\Mpl^{2}k^{2}+24a^{2}V_{,y}X^{2}}>0\,,
\end{equation}
implying
\begin{equation}
V_{,y}>0\,,
\end{equation}
which coincides to one of the conditions already found for the scalar
modes.

\subsection{Tensor modes}

The tensor modes are not affected by the presence of the 3-form, as
this latter one is minimally coupled to gravity and it does not possess
tensor degrees of freedom. To show this more in detail, we choose
the tensor perturbations as $\delta g_{ij}=h_{ij}^{T}=h_{+}e_{ij}^{+}+h_{\times}e_{ij}^{\times}$,
where both the symmetric tensors $e_{ij}$ are transverse and traceless.
We also impose the normalization condition, $e_{ij}(\bm{k})e_{ij}(-\bm{k})^{*}=1$,
for each polarization, whereas $e_{ij}^{+}(\bm{k})e_{ij}^{\times}(-\bm{k})^{*}=0$.
Therefore the second order action can be written as
\begin{equation}
S_{T}=\sum_{\lambda=+,\times}\int dt\, d^{3}x\, a^{3}\,\frac{\Mpl^{2}}{8}\left[\dot{h}_{\lambda}^{2}-\frac{1}{a^{2}}(\partial h_{\lambda})^{2}\right],
\end{equation}
so that no stability condition comes from the tensor sector.

\section{Suitable form of potentials for 3-form inflation\label{sec:potential}}

According to the previous section, one of the no-ghost condition can
be written as $V_{,y}>0$, where $y=6X^{2}\geq0$. The existence of
ghosts in the model depends on the shape of three form potential,
but not on the sign of $X$ (as $y\propto X^{2}$). In fact, in order
to search for the form of potentials, which makes the 3-form field
ghost-free and without Laplacian instabilities ($c_{X}^{2}\geq0$),
we need to study more in detail the evolution of $y$ (or, equivalently,
$X$). It is convenient for qualitative analysis to change variables
to dimensionless variables and define some quantities. In the first
part of this section we will define some quantities and use some dimensionless
variables as found in \cite{Koivisto:2009fb}. From the Friedmann
equation, we have
\begin{equation}
\dot{H}=-\frac{1}{\Mpl^{2}}\, V_{,y}\, y=-\frac{1}{2\Mpl^{2}}\, V_{,X}X\,,\label{eq:supaX}
\end{equation}
so that the 3-form field can play the role of a slow-rolling inflaton
if $V_{,X}X/\Mpl^{2}\ll H^{2}$. Substituting the above Eq.~(\ref{eq:supaX})
into the evolution Eq.~(\ref{eq:eqX}), we get
\begin{equation}
\ddot{X}+3H\dot{X}+V_{{\rm eff},X}=0\,,\label{eq:ddotx}
\end{equation}
where
\begin{equation}
V_{{\rm eff},X}=\frac{dV_{{\rm eff}}}{dX}=V_{,X}\left(1-\frac{3}{2}\,\frac{X^{2}}{\Mpl^{2}}\right),\label{eq:vfx}
\end{equation}
so that the effective potential is given by
\begin{equation}
V_{{\rm eff}}(X)=\int^{X}d\xi\, V_{,\xi}\left(1-\frac{3}{2}\,\frac{\xi^{2}}{\Mpl^{2}}\right).
\end{equation}

On using the dimensionless variables
\begin{equation}
x\equiv\frac{X}{\Mpl}\,,\qquad{\rm and}\qquad w\equiv\frac{3x+x'}{\sqrt{6}}\,,
\end{equation}
where a prime denotes a derivative with respect to $N=\ln a$, Eq.~(\ref{eq:ddotx})
can be written in the autonomous form as
\begin{eqnarray}
x' & = & 3\left[\sqrt{\frac{2}{3}}w-x\right]\,,\label{xp}\\
w' & = & \frac{3}{2}\,\lambda(x)\,(1-w^{2})\left(x\, w-\sqrt{\frac{2}{3}}\right),\label{wp}
\end{eqnarray}
where we have introduced the function
\begin{equation}
\lambda\equiv\frac{V_{,x}}{V}\,.
\end{equation}
In these variables the slow roll parameter can be written as
\begin{equation}
\epsilon\equiv-\frac{\dot{H}}{H^{2}}=\frac{3}{2}\,\lambda\,(1-w^{2})\, x\,.
\end{equation}
The accelerating expansion of the universe is acquired by demanding
$\epsilon\ll1$. From this parameter, one can see that the kinetic
term does not necessarily need to be small compared to the potential
term, as for the standard picture of the inflaton scalar field. Conversely,
it requires that $w^{2}\approx1$ when $x\,\lambda(x)\sim O(1)$.
In order to have inflation, one needs one more requirement to guarantee
that the accelerating expansion is long enough. We introduce a parameter
to characterize this behavior as
\begin{equation}
\eta\equiv\frac{\epsilon'}{\epsilon}-2\epsilon=(1+c_{X}^{2})\,\frac{x'}{x}\,,\label{eta}
\end{equation}
where the inflationary period requires that $|\eta|\ll1$. Since $c_{X}^{2}>0$,
$|\eta|$ will be small if $x'/x$ is small, that is $x$ needs to
be in a slow-roll regime. From Eq.~(\ref{xp}), it implies that $|\eta|\ll1$
will be satisfied if $x\simeq\sqrt{2/3}w\simeq\pm\sqrt{2/3}$, where
we have imposed also the first slow-roll condition $|\epsilon|\ll1$
when $x\,\lambda(x)\sim O(1)$.

We note that, as a consequence of the definition of $w$, we have
$Y/(\Mpl H)=\sqrt{6}\, w$. Therefore, the Friedmann equation (\ref{eq:fried})
implies
\begin{equation}
1=\rho_{X}/(3\Mpl^{2}H^{2})=w^{2}+V/(3\Mpl^{2}H^{2})\,,
\end{equation}
so that, if $V\geq0$, then $0\leq w^{2}\leq1$.

As we have already said, the field will slow-roll when it reaches
the points $P\equiv(x,w)=(\pm\sqrt{2/3},\pm1)$ in phase-space because
these points are (de Sitter) fixed points (unless $\lambda(x)$ is
not finite at these points). There might be other fixed points, $M$,
which correspond to the points where $\lambda$ vanishes, that is
$M\equiv(x,w)=(\bar{x},\sqrt{3/2}\bar{x})$, where $\lambda(x=\bar{x})=0$.
It can be seen from Eq.~(\ref{eq:vfx}) that the points $P$ and
$M$ are the values of $X$ which correspond to the extrema of the
effective potential.

\subsection{Stability of the fixed points}

Let us start by studying the stability of the fixed point $P$. By
choosing $x=\pm\sqrt{2/3}+\delta x$, and $w=\pm1+\delta w$, we can
linearize the equations of motion with respect to the small quantities
$\delta x$ and $\delta w$, and we find
\begin{eqnarray}
\delta x' & = & \sqrt{6}\,\delta w-3\delta x\,,\\
\delta w' & = & 0\,,
\end{eqnarray}
with solutions $\delta w=b_{1}={\rm constant}$, $\delta x=\sqrt{2/3}b_{1}+b_{2}e^{-3N}$,
and $b_{1,2}$ are (small) initial conditions.
From the autonomous system in Eq.~(\ref{xp}) and Eq.~(\ref{wp}), we find that the eigenvalues for this fixed point are (-3, 0).
The fact that one of the eigenvalues is zero,
implies that, at linear order, we cannot deduce whether the fixed
point is stable or not. In order to check the stability of this fixed
point, one needs to study also the second-order solution. For the
second order perturbation, it is convenient to parametrize the perturbation
variables in such that $\delta x'=\delta w'=0$. This corresponds
to choosing the perturbation variables along the eigenvector which
has zero eigenvalue. By using $\delta x'=0$, one finds that $\delta x=\sqrt{\frac{2}{3}}\delta w$.

Therefore, by keeping the perturbations up to second order, we find
\begin{eqnarray}
\delta w'=-2\sqrt{6}\lambda(\pm\sqrt{2/3})\,\delta w^{2},
\end{eqnarray}
which can be solved as
\begin{eqnarray}
\delta w=\frac{\delta w_{0}}{1\,+\,2\sqrt{6}\lambda(\pm\sqrt{2/3})\delta w_{0}N}
\end{eqnarray}
where $\delta w_{0}=\delta w(N=0)$. To ensure the stability of the
perturbation, one requires a condition
\begin{eqnarray}
\lambda(\pm\sqrt{2/3})\delta w_{0}\,>\,0.
\end{eqnarray}
Since we have $-1\leq w\leq1$, $\delta w_{0}$ must be negative at
fixed point ($+\sqrt{2/3},+1$) and $\delta w_{0}$ must be positive
at fixed point ($-\sqrt{2/3},-1$). Therefore, the condition above
becomes
\begin{eqnarray}
\lambda(+\sqrt{2/3})=\left.\frac{V_{,x}}{V}\right|_{x=+\sqrt{2/3}} & < & 0\,,\\
\lambda(-\sqrt{2/3})=\left.\frac{V_{,x}}{V}\right|_{x=-\sqrt{2/3}} & > & 0.
\end{eqnarray}
For viable three-form model, which $xV_{,x}>0$ and $V>0$, these
conditions show that, at second order, the fixed point is unstable.
This second order perturbation analysis is equivalent to one in \cite{Koivisto:2009fb}
and also agrees with the numerical calculation in \cite{Ngampitipan:2011se}.
We note that there is another method to find the stability of the
fixed point which has zero eigenvalue as shown in \cite{Boehmer:2011tp}.

The fact that this instability appears at second order means that
the instability will in general evolve slowly. This instability will
make inflation end eventually. Now we have one more condition for
viable inflationary model from three-form field which is the point
$P$ must be unstable. This requirement is also compatible with ghost-free
condition. Furthermore, one can rule out some potential forms by using
these condition. By considering various potentials which have been
investigated in \cite{Koivisto:2009fb}, one finds that the model
with Mexican-hat potential, $V=V_{0}\,(x^{2}-c^{2})$ is plagued by
a ghost. In the case $c>\sqrt{2/3}$, there is a ghost and the point
$P$ is stable. For the case $c<\sqrt{2/3}$, even though the point
$P$ is not stable, the field $x$ will evolve to oscillate around
$x=c$ at the end of inflation and then a ghost eventually appears
when $x<c$. We note that, for the shift-potential $V=V_{0}\,(x^{2}-c^{2})+k$
where $k$ is positive constant, ghost will appear since slope of
the potential is the same. For the case $c<\sqrt{2/3}$, the field
$x$ may not cross $x=c$ if this point is stable fixed point. However,
the inflation will occur again since the field slowly move to this
fixed point.

To obtain the suitable potential form for inflation, there must contain
the oscillating phase which provides the possibility for reheating
period. To avoid a ghost during oscillating phase, the viable potential
form must have only one minimum locating at $x=0$ which is not stable
fixed point. Therefore, we will find the property of this fixed point
next.

As for the fixed point $M=(x,w)=(\bar{x},\sqrt{3/2}\bar{x})$, where
$\lambda(\bar{x})=0$, then by choosing $x=\bar{x}+\delta x$, and
$w=\sqrt{3/2}\bar{x}+\delta w$, we find the linearized equations
\begin{eqnarray}
\delta x' & = & \sqrt{6}\delta w-3\delta x\,,\\
\delta w' & = & -\frac{\sqrt{6}}{8}\,(2-3\bar{x}^{2})^{2}\bar{\Gamma}\delta x\,,
\end{eqnarray}
where
\begin{equation}
\bar{\Gamma}=\left.\frac{V_{,xx}}{V}-\left(\frac{V_{,x}}{V}\right)^{2}\right|_{x=\bar{x}}=\left.\frac{V_{,xx}}{V}\right|_{x=\bar{x}}.
\end{equation}
The solution leads to
\begin{equation}
\delta x=d_{1}e^{-N\,(3+\gamma)/2}+d_{2}e^{-N\,(3-\gamma)/2}\,,
\end{equation}
where
\begin{equation}
\gamma=\sqrt{9-3\bar{\Gamma}\,(3\bar{x}^{2}-2)^{2}}\,.
\end{equation}
An instability will appear if $\gamma>3$, or $\bar{\Gamma}<0$. For
the fixed point which $\bar{x}=0$, one found that $\bar{\Gamma}>0$
for the positive even potential. Thus this fixed point is always stable.
One of the way to get the oscillating phase is that $\lambda(x=0)$
must be not finite. Therefore, one requires more condition of the
potential form that $V$ must vanish at $x=0$, $V(x=0)\,=\,0$. This
requirement will rule out the potential forms which have been investigated
in \cite{Koivisto:2009fb} such that $V=V_{0}\, e^{\alpha x^{2}}$,
$V=V_{0}\,(x^{2}+\alpha)$ and $V=V_{0}\,(x^{4}+\alpha)$ where $\alpha$
is a positive constant. Now we can summarize that the viable potential
forms which have been investigated in \cite{Koivisto:2009fb} are
only $V=V_{0}\, x^{2}$ and $V=V_{0}\, x^{4}$.

Generally, a power law potential of the form $V\propto y^{p}\,=\, x^{2p}$
will be suitable potential form for inflationary model from three-form.
However, on choosing a power law potential of the form $V\propto y^{p}$,
we immediately notice that $Q\propto y^{p-1}$, which in general vanishes
(for $p>1$) or diverges (for $p<1$) as $y\to0$, unless $p=1$.
Since we will focus on values $p\geq1$, most of the potential will
allow the field to cross this value ($y=0$), so that $Q$ will vanish
in the origin. This property represents a problem, in general, as
this means that, at that point, the second order Lagrangian vanishes
(as $c_{X}^{2}$ remains finite for $V\propto y^{p}$), and the theory
becomes strongly coupled, i.e.\ higher order corrections become dominant.
In fact, the metric curvature perturbation, in order that perturbation
theory makes sense, needs to be smaller than unity for all dynamics.
Therefore, in the limit that $Q\to0$, the whole action, if $c_{X}^{2}$
remains finite, will tend to vanish. It is worthy to recall that at
the point $V_{,y}=0$ where $Q$ vanishes, the chosen gauge is in
general well defined, unless at that point we have that $Y=\dot{X}+3HX=0$,
which for $X=0$, it implies $\dot{X}=0$. But $X=0=\dot{X}$
is not a point which is reached by the dynamics in a finite interval
of time, in general.

In order to avoid this possible strong-coupling issue we propose the
following generalized power law potential%
\footnote{In general, we can choose a larger class of potentials given as $V(y)=V_{0}\,\bigl(c\, y+\sum_{i}c_{i}\, y^{p_{i}}\bigr)$,
where $c>0$, $c_{i}\geq0$, and $p_{i}\geq1$.%
},
\begin{equation}
V(x^{2})=V_{0}\,[\,(x^{2})^{p}+bx^{2}]\,,\label{eq:pogen}
\end{equation}
where $p$ is a constant which can be, as for now, positive or negative,
whereas $b>0$. Note that for the potential form, $V=V_{0}\, x^{2}$,
it has been investigated in detail in \cite{Koivisto:2009ew}. We
can also modify the Gaussian potential in order to satisfy the conditions
as
\begin{equation}
V(x^{2})=V_{0}\,[e^{\nu x^{2}}-1]\,,\label{eq:gaussgen}
\end{equation}
where $\nu$ is a positive constant parameter. We will investigate
the properties of these potential forms in the next subsection.

\subsection{Power-law potential}

\begin{figure}
\includegraphics[width=7cm]{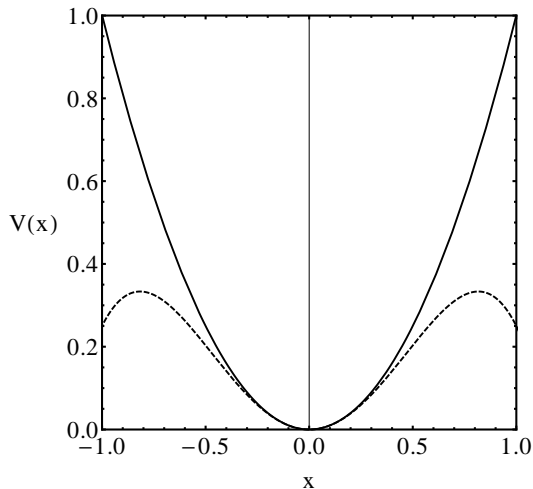} \caption{\label{fig:5} The potential $V(x)\propto x^{2}$ is shown here. In
the figure, the bare potential and the effective potential are represented
by a solid and dashed line respectively.}
\end{figure}

We now investigate cosmological behavior for the potential
\begin{equation}
V(x^{2})=V_{0}\,[(x^{2})^{p}+b\, x^{2}]\,,\label{eq:po}
\end{equation}
where $p$ is a constant which can be, as for now, positive or negative,
whereas $b>0$. For this form of the potential, we have that the no-ghost
condition
\begin{equation}
\frac{V_{,x}}{x}=2V_{0}[p\,(x^{2})^{p-1}+b]>0\,,\label{eq:gh-po}
\end{equation}
is always positive for $p\geq0$, and also finite for $p\geq1$. For
this reason, from now on, we will consider only the case $p\geq1$.
On the other hand, since
\begin{equation}
c_{X}^{2}=\frac{(2p-1)\, p\,(x^{2})^{p-1}+b}{p\,(x^{2})^{p-1}+b}\,,
\end{equation}
$c_{X}^{2}$ will be always positive and finite for $p\geq1$. The
bottom line is that for $p\geq1$ the model is free from instabilities
for any real value of $x$, that is for any dynamics. It should be
noticed that for the value $p=1$, the potential reduces to a quadratic
power law potential. On the de Sitter fixed point $P$, we have
\begin{equation}
c_{X}^{2}(x=\pm\sqrt{2/3})=\frac{(2p-1)\, p\,(2/3)^{p-1}+b}{p\,(2/3)^{p-1}+b}\geq1\,,
\end{equation}
and the inequality holds for $p\geq1$, and $b>0$.

Because of Eq.~(\ref{eq:po}), it can be shown that
\begin{equation}
V_{{\rm eff},x}=2V_{0}\left(1-\frac{3}{2}\, x^{2}\right)\left[p\, x^{2p-1}+b\, x\right]\,,
\end{equation}
and
\begin{equation}
V_{{\rm eff}}=V_{0}\left[\frac{x^{2p}}{2(p+1)}[2+p(2-3\, x^{2})]+bx^{2}\left(1-\frac{3}{4}\, x^{2}\right)\right],
\end{equation}
so that the extremum points of this potential occur at $x=\pm\sqrt{2/3}$,
and $x=0$. We also notice that
\begin{equation}
\lambda(x)\, x=\frac{2\,[p\,(x^{2})^{p-1}+b]}{[(x^{2})^{p-1}+b]}\,,
\end{equation}
which, for $p\geq1$, is positive and finite for all $x$. Furthermore
$\lim_{x\to0}\lambda\, x=2$.

\begin{figure}
\includegraphics[height=7cm]{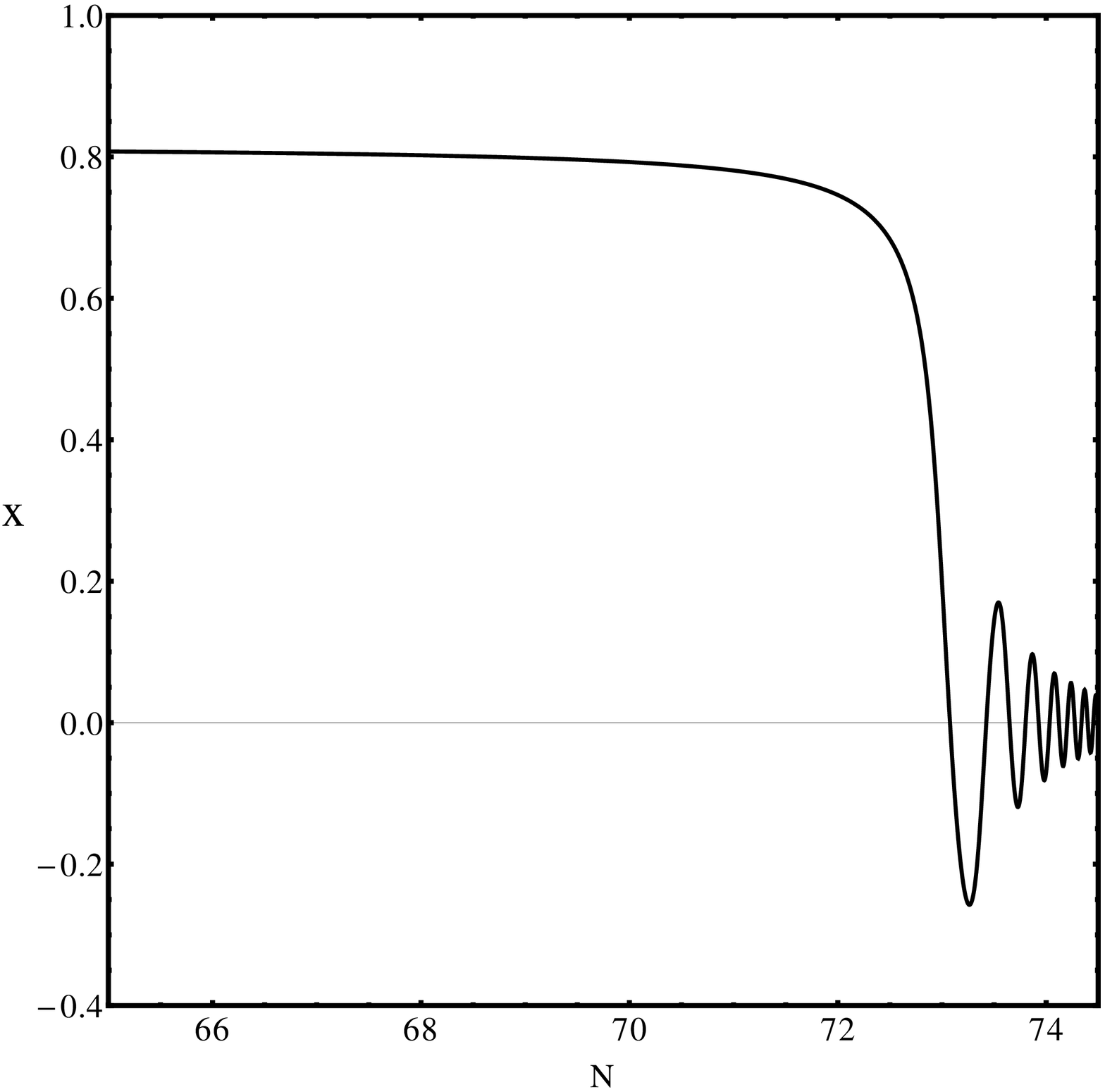}\,\,\,\,\,\,\includegraphics[height=7cm]{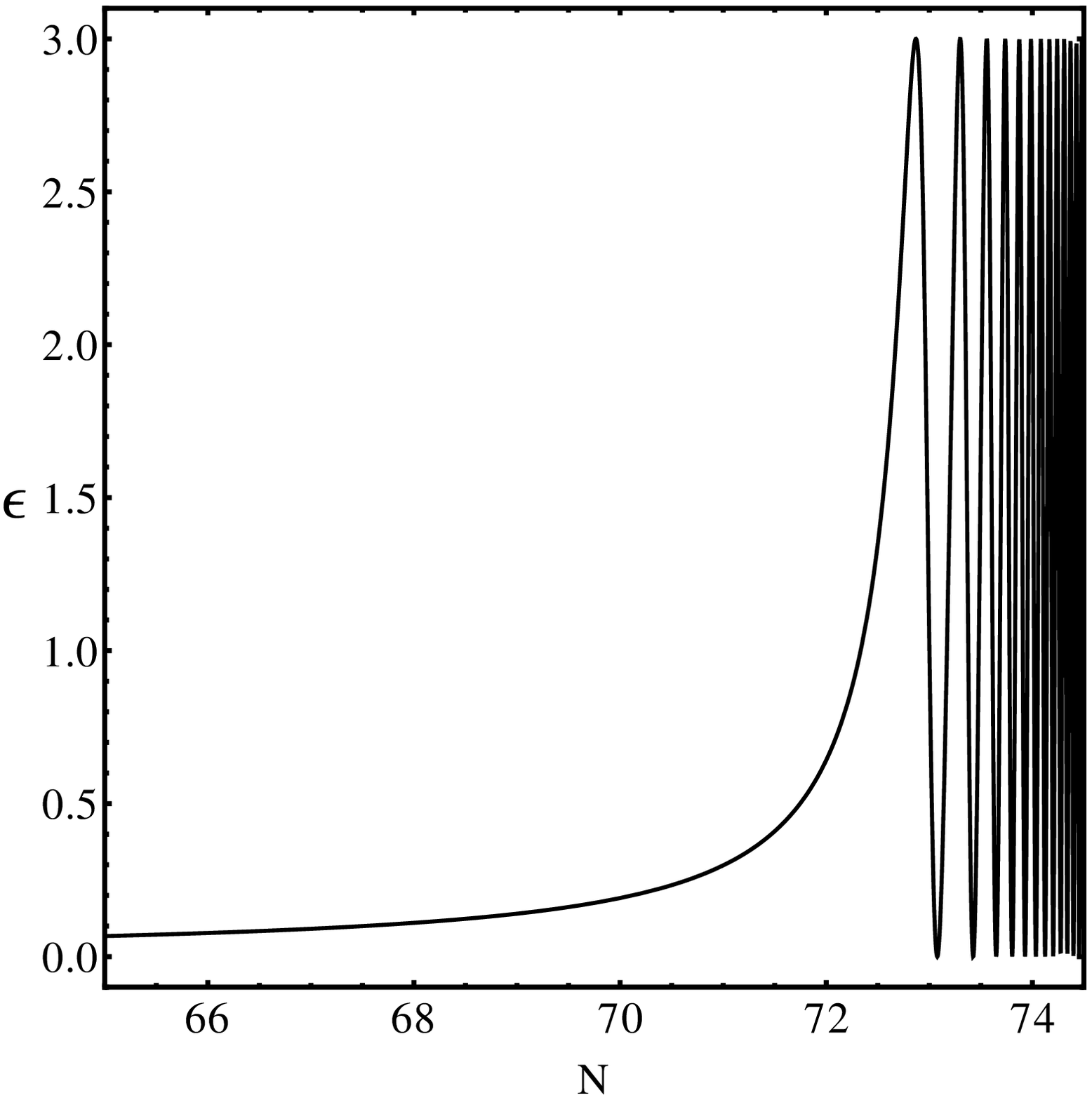}
\caption{\label{fig:pl-1} The evolution of $x$ and $\epsilon$ for the potential
$V=V_{0}(1+b)\, x^{2}$, and $p=1$. We chose initial conditions $w(0)=0.99$,
and $x(0)=3\sqrt{2/3}\, w(0)$. When the field $x$ start to oscillate
around the minimum, the parameter $\epsilon$ will start to oscillate
around $\epsilon=3/2$. Thus the inflation will end when $\epsilon\sim1$
corresponding to $N\sim72$. }
\end{figure}

For illustration, we consider here the simplest case $p=1$ whereas,
in the next section, we will describe in more detail the case $p=2$
(quartic potential). The potential and effective potential for the
$p=2$ case are plotted in Figure \ref{fig:5}. The field that starts
at $|x|>\sqrt{2/3}$ with $|w|\sim1$ is able to drive long-enough
inflation. However, this time, the field rolls down the potential
from the points A or B and then oscillates around the minimum of the
potential without ghosts ($V_{,y}=V_{0}(1+b)/6>0$) or Laplacian instabilities
($c_{X}^{2}=1$). We also use a direct numerical integration to confirm
both the slow roll and the oscillatory regimes as shown in Figure
\ref{fig:pl-1}. From the evolution of $\epsilon$ in the right panel,
inflation ends at $N\sim72$ corresponding to $\epsilon\sim1$.

\subsection{Quartic potential}

\begin{figure}
\includegraphics[height=7cm]{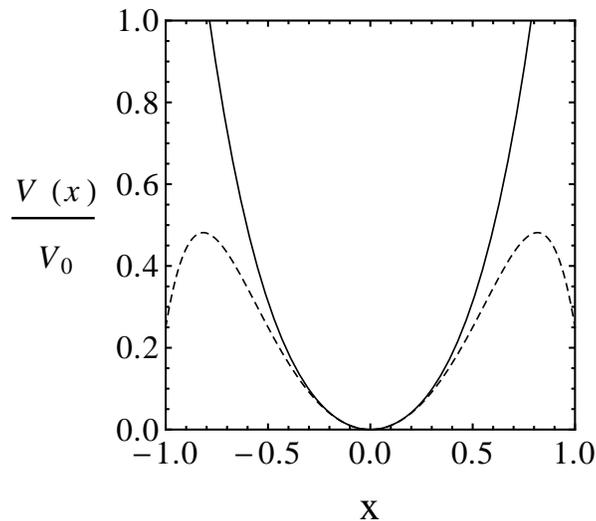} \caption{\label{fig:q-0} The potential $V=V_{0}(x^{4}+bx^{2})$, with $b=1$
(the continuous and black curve), together with the effective potential
$V_{{\rm eff}}$ (represented by the dashed curve).}
\end{figure}

We study here a particular case of the power law potential introduced
in the previous section, namely
\begin{equation}
V=V_{0}(x^{4}+b\, x^{2})\,,
\end{equation}
and we plot it (together with its effective potential) in Figure \ref{fig:q-0}.

\begin{figure}
\includegraphics[height=7cm]{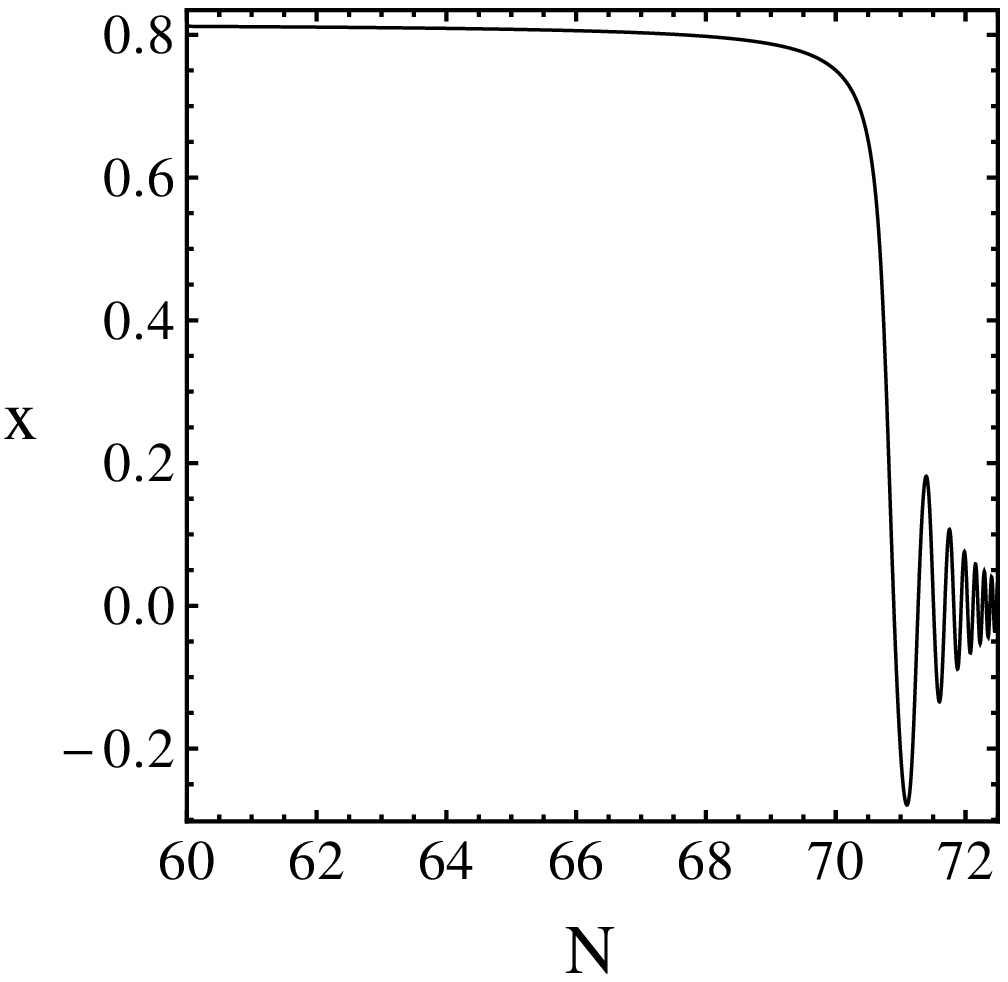}\,\quad{}\,\enskip{}\quad{}\,\includegraphics[height=7cm]{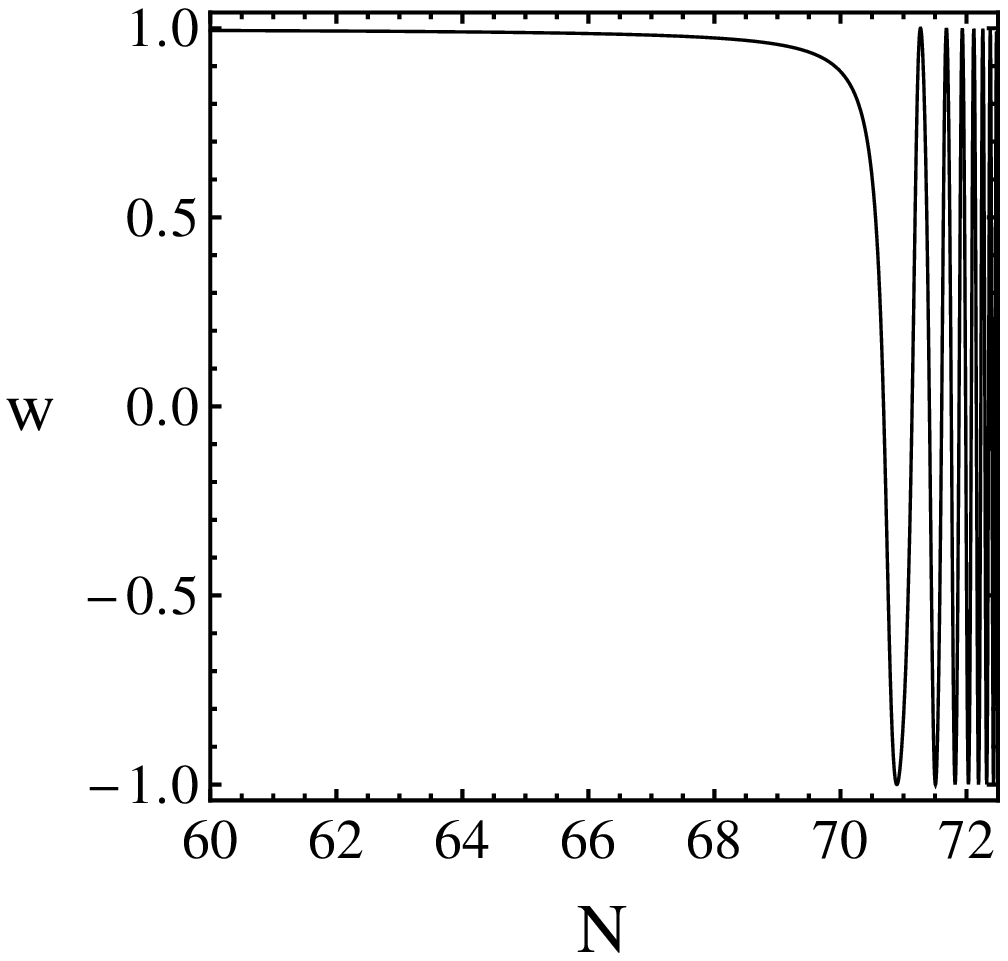}
\caption{\label{fig:q-1} The evolution of $x$ and $w$ for the potential
$V=V_{0}(x^{4}+bx^{2})$, and $b=1$. We chose initial conditions
$w(0)\approx0.9991$, and $x(0)\approx0.8158$.}
\end{figure}

In this case, we also have

\begin{equation}
\frac{3\Mpl^{2}H^{2}}{V_{0}}=\frac{x^{4}+bx^{2}}{1-w^{2}}\,,
\end{equation}
so that, by introducing a dimensionless cosmic time $t$, we can write
\begin{equation}
\frac{dx}{dt}=\frac{\Mpl H}{\sqrt{V_{0}/3}}\,\frac{dx}{dN}=\sqrt{\frac{x^{4}+bx^{2}}{1-w^{2}}}\,\frac{dx}{dN}\,.
\end{equation}
In Figure \ref{fig:q-1}, we show the evolution for both $x$, and
$w$. The field slow-rolls until an oscillatory regime starts, making
inflation end.

\begin{figure}
\includegraphics[height=7cm]{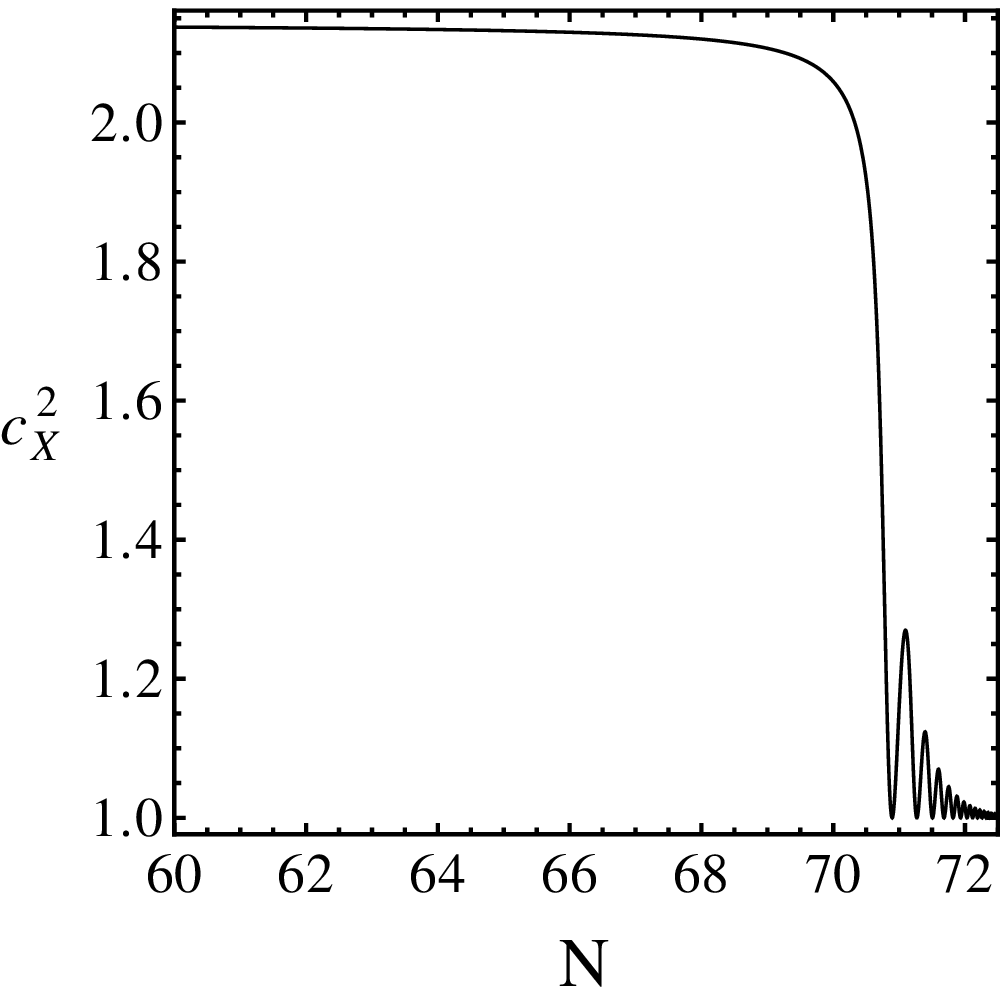} \caption{\label{fig:q-2} The evolution of $c_{X}^{2}$ for the potential $V=V_{0}(x^{4}+bx^{2})$,
and $b=1$. The model does not possess any instability.}
\end{figure}

During the slow-roll regime the propagation speed takes the value
$c_{X}^{2}\approx\frac{3(4+b)}{4+3b}$, whereas, as the solution starts
oscillating, $c_{X}^{2}\to1$. This behavior is confirmed in Figure
\ref{fig:q-2}.

\begin{figure}
\includegraphics[height=7cm]{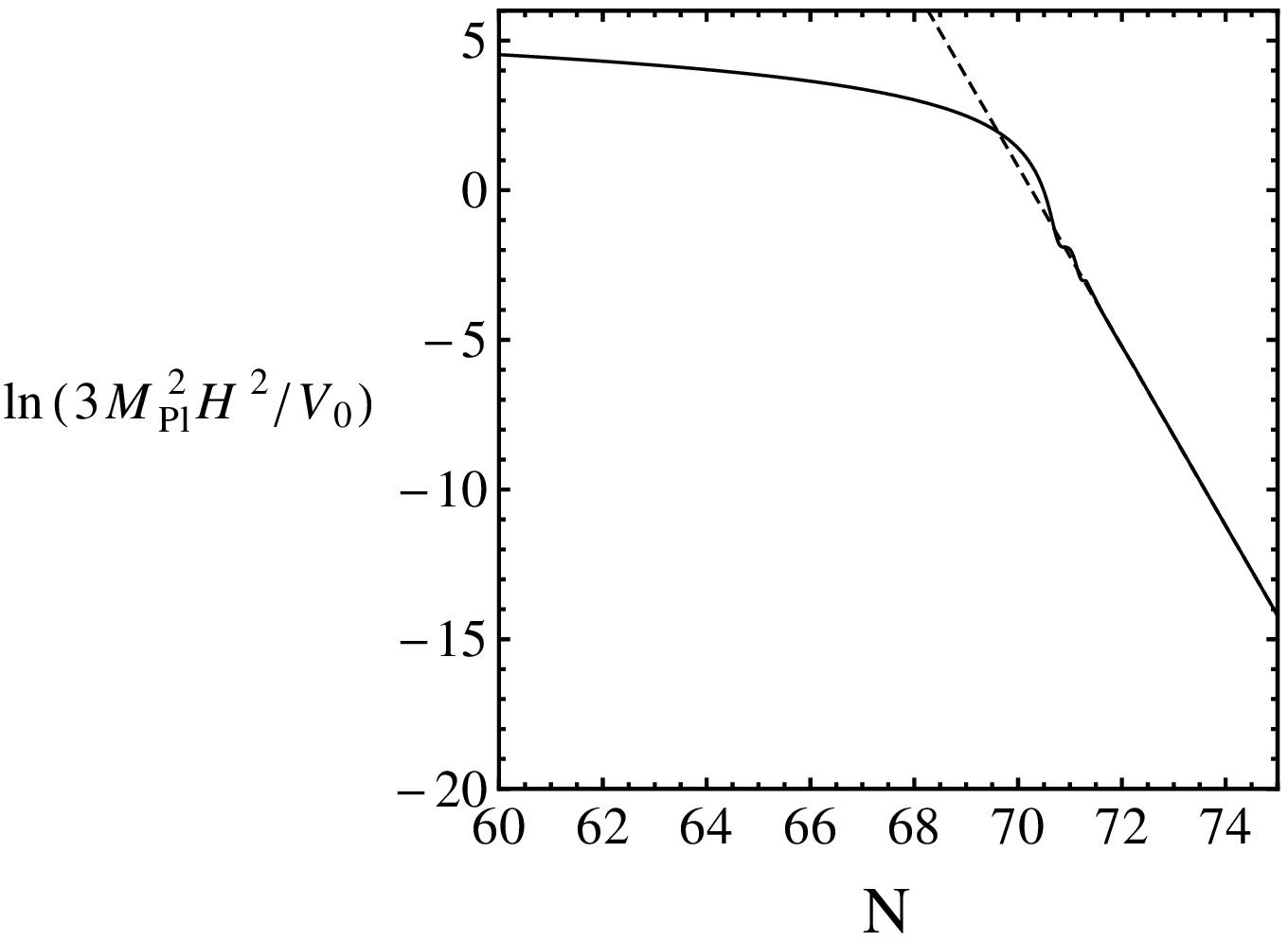} \caption{\label{fig:q-3} The evolution of $3\Mpl^{2}H^{2}/V_{0}$ for the
potential $V=V_{0}(x^{4}+bx^{2})$, and $b=1$ (continuous black curve).
This figure shows that after inflation ends (around $N\simeq70$),
the universe enters a matter dominated epoch, as the curve approaches
a dashed line, which represents the line $\ln(3\Mpl^{2}H^{2}/V_{0})=-3N+{\rm constant}$.
This means that after inflation $H^{2}\propto e^{-3N}\propto a^{-3}$.}
\end{figure}

Finally, we show in Figure \ref{fig:q-3} that after inflation ends,
there is an oscillatory regime which mimics a dust dominated universe
as we have $H^{2}\propto a^{-3}\propto e^{-3N}$. This behavior is
similar to the standard single-field inflationary models, and this
is not surprising, because as $x\to0$, we find $V_{{\rm eff}}\approx b\, x^{2}$;
so that the equation of motion for the field, Eq.~(\ref{eq:eqX}),
reduces to
\begin{equation}
\ddot{x}+3H\dot{x}\approx-bx\,,\qquad{\rm for\ }|x|\ll1\,,
\end{equation}
which exactly matches the equation of motion for standard inflation
in the presence of a quadratic inflaton potential. In other words,
the dynamics of the 3-form, for $x\to0$, tends to be more and more
identical to the dynamics of a single scalar field oscillating around
the minimum of a quadratic potential.

\begin{figure}
\includegraphics[height=7cm]{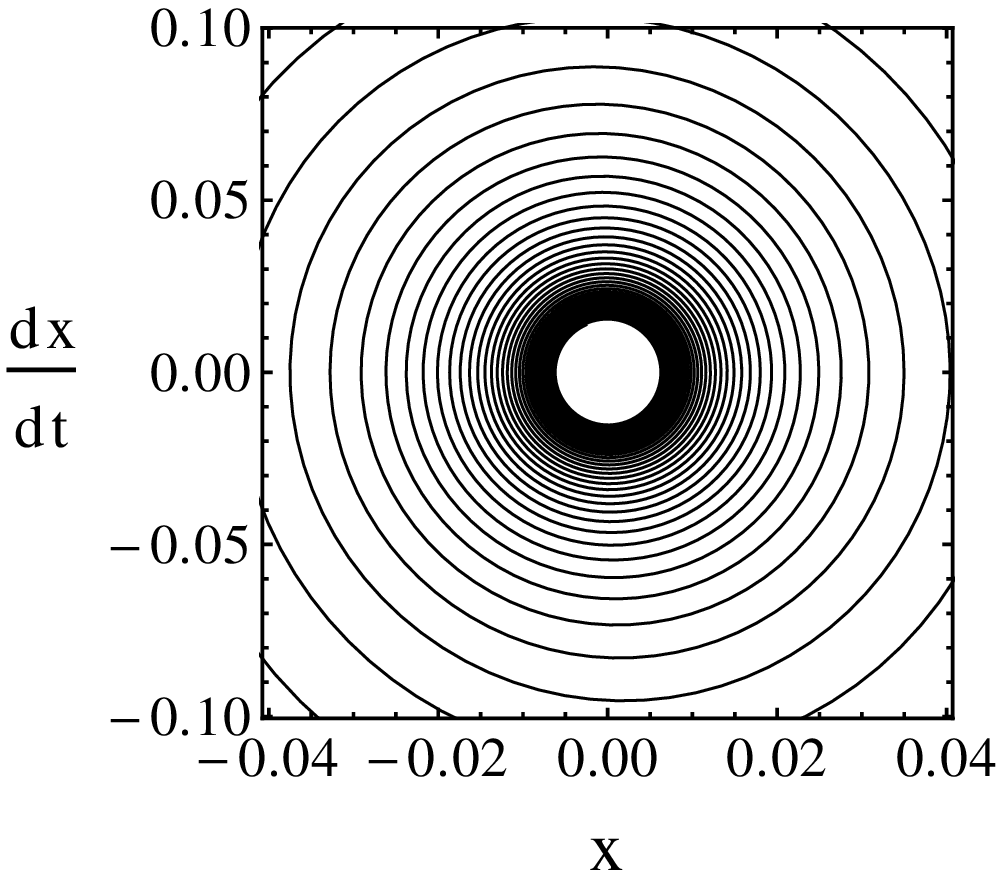}\,\quad{}\,\enskip{}\quad{}\,\includegraphics[height=7cm]{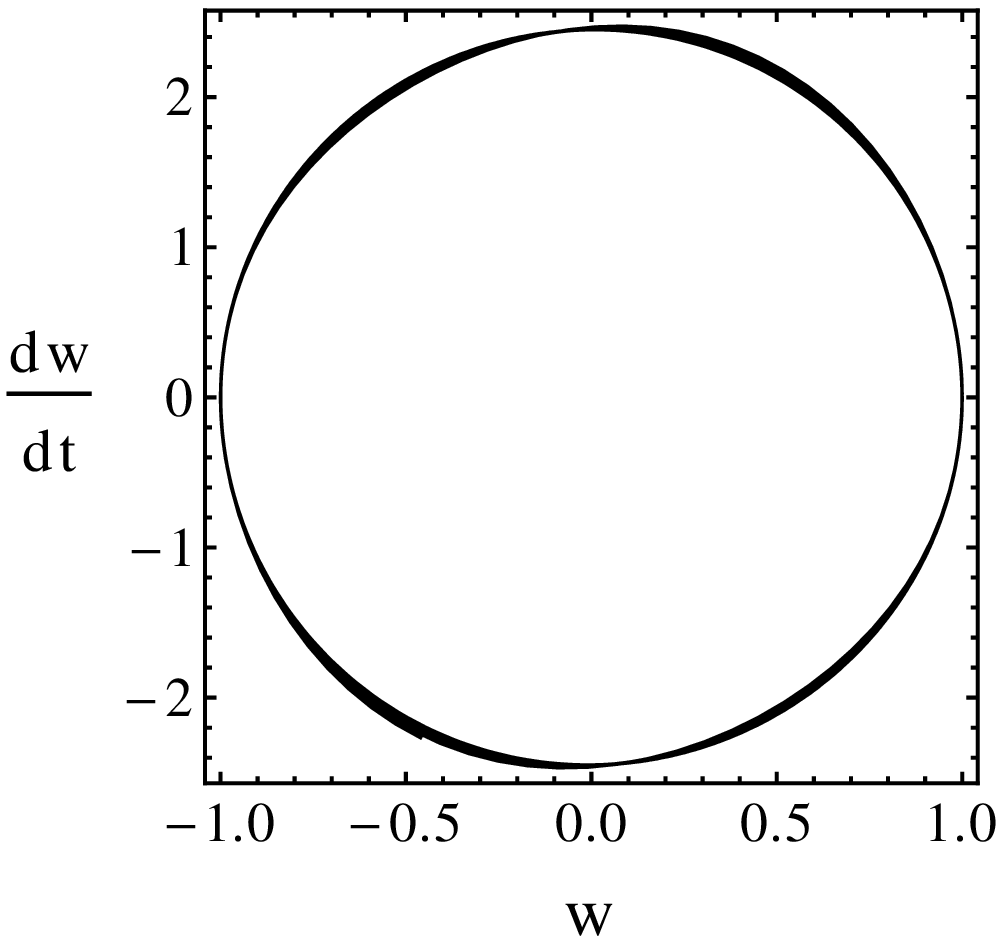}
\caption{\label{fig:q-4} The evolution of $dx/dt$ as a function of $x$,
and $dw/dt$ as a function of $w$ for the potential $V=V_{0}(x^{4}+bx^{2})$,
and $b=1$ during the oscillatory regime. We chose initial conditions
so that at $t=1$, the values of $x$ and $w$ correspond respectively
to $x(N=72)\approx0.0644$, and $w(N=72)\approx-0.46$ of Figure \ref{fig:q-1}.
We stop the integration at $t=100$. This figure shows that $x$ spiralizes,
whereas $w$ keeps on oscillating during the matter dominated regime.}
\end{figure}

After inflation ends, during the oscillatory regime, in Figure \ref{fig:q-1},
we see that $x\to0$, whereas $w$ oscillates between $-1$ and 1.
Furthermore, we also find that $dx/dt\to0$ together with $x$, whereas
$dw/dt$ keeps oscillating, remaining finite, as shown in Figure \ref{fig:q-4}.

\subsection{Gaussian potential}

\begin{figure}
\includegraphics[height=7cm]{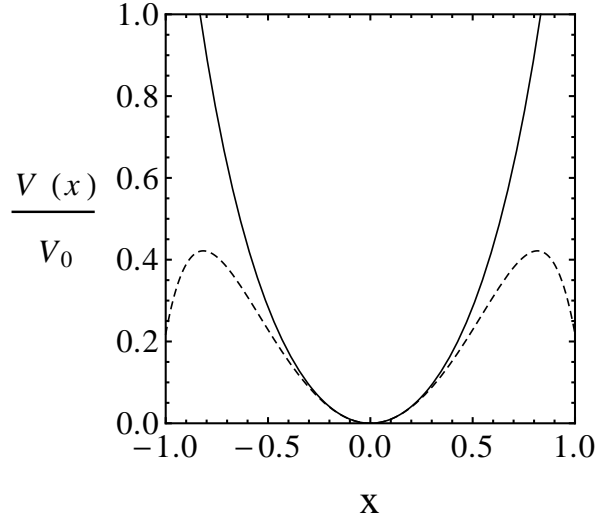} \caption{\label{fig:gs-1} The potential $V(x)=V_{0}\,(e^{\nu\, x^{2}}-1)$
is represented by a solid line (for $\nu=1$), whereas the effective
potential is represented by a dashed line. }
\end{figure}

We now consider the exponential potential
\begin{equation}
V=V_{0}(e^{\nu y/6}-1)=V_{0}(e^{\nu x^{2}}-1)\,,\label{Gauss}
\end{equation}
where $\nu$ is a constant parameter which can be positive or negative.
For this form of potential, we have
\begin{equation}
\frac{V_{,x}}{x}=2\nu V_{0}\, e^{(\nu x^{2})}\,,
\end{equation}
so that the ghost will not exist if $\nu$ is positive. The speed
of propagation in this case is given by
\begin{equation}
c_{X}^{2}=1+2\nu x^{2}\,.
\end{equation}
This implies that if the ghost does not exist, $c_{X}^{2}$ is always
positive. Substituting Eq.~(\ref{Gauss}) into Eq.~(\ref{eq:vfx}),
one gets
\begin{equation}
V_{{\rm eff},x}=2\nu xV_{0}e^{\nu x^{2}}\left(1-\frac{3}{2}\, x^{2}\right)\,,\label{vfx-g}
\end{equation}
or
\begin{equation}
V_{{\rm eff}}=-\frac{V_{0}}{2\nu}\left\{ 3+2\nu+e^{\nu x^{2}}[(3x^{2}-2)\nu-3]\right\} ,
\end{equation}
and we plot it, together with the bare potential, in Figure \ref{fig:gs-1}.
We also notice that as $x\to0$, then $V_{{\rm eff}}\simeq V_{0}\nu x^{2}$,
so that we expect an oscillatory regime to take place, ending inflation.

\begin{figure}
\includegraphics[height=7cm]{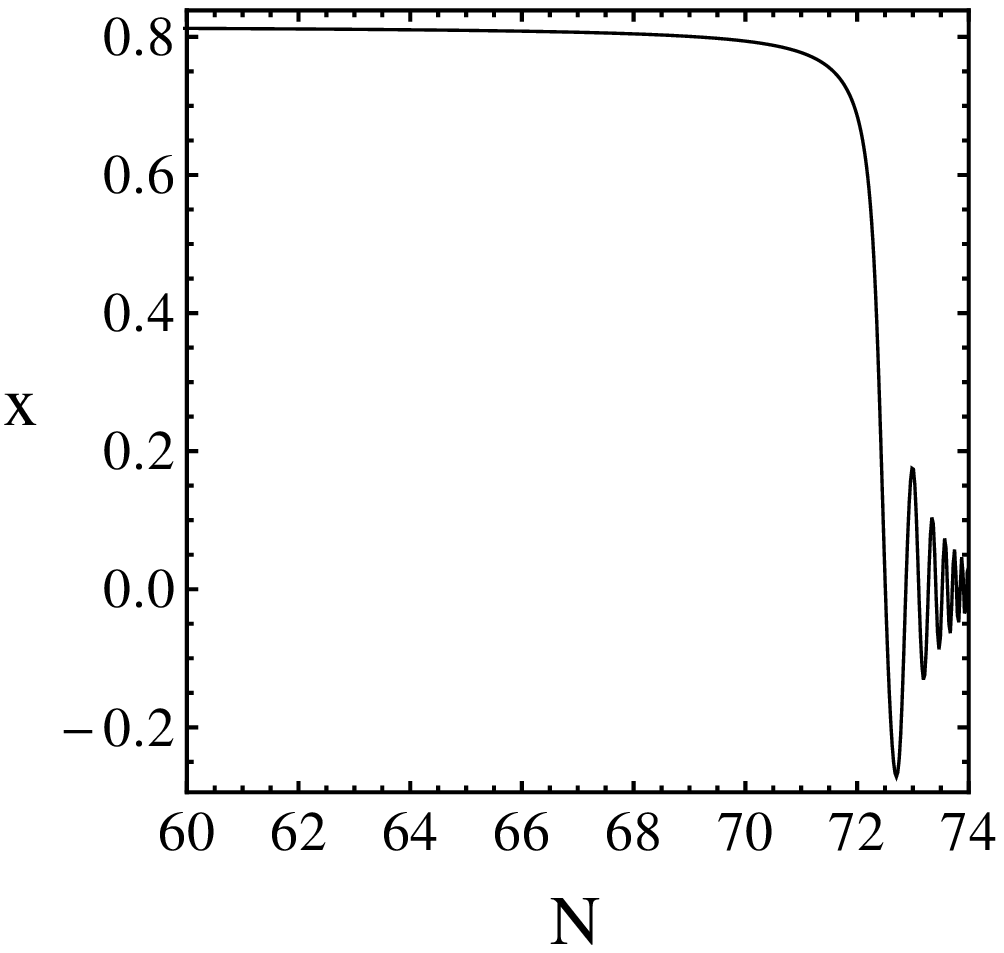}\,\,\,\,\,\,\includegraphics[height=7cm]{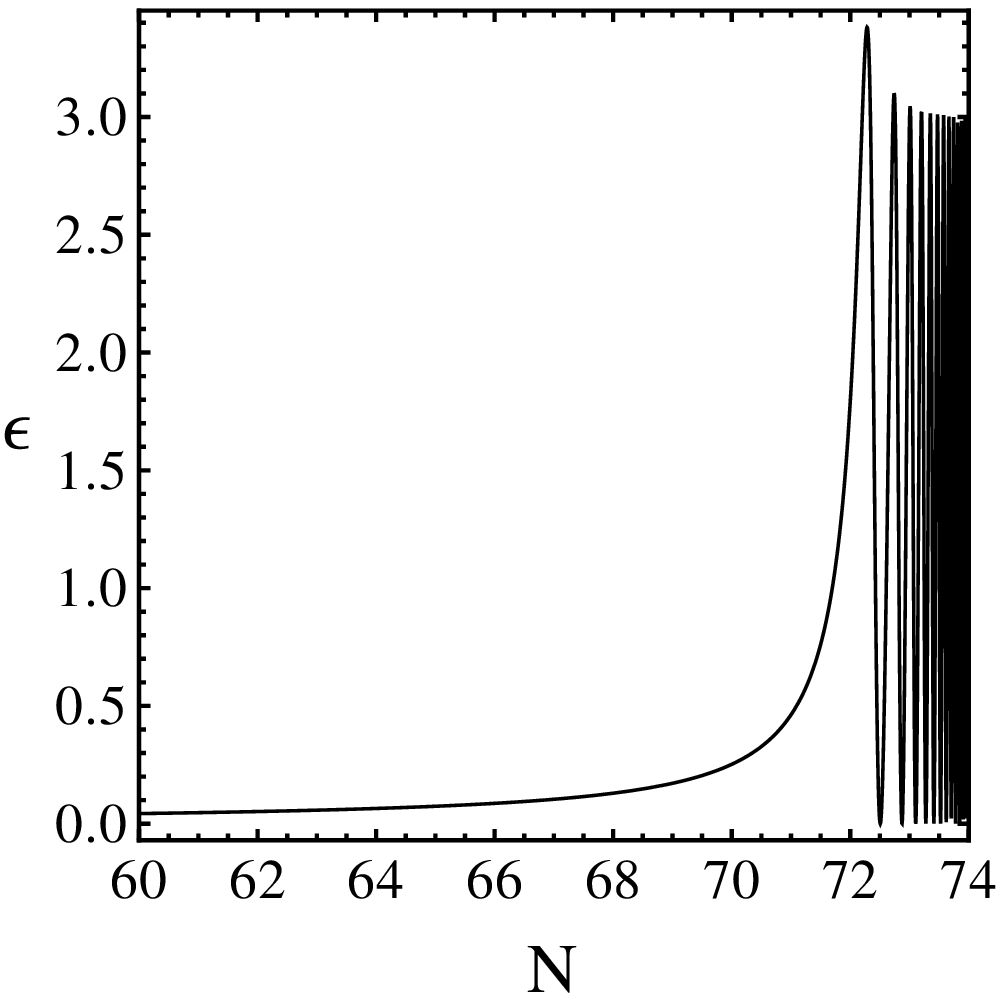}
\caption{\label{fig:gs-2} The evolution for $x$ (left panel) and $\epsilon$
(right panel) for the potential $V_{0}\,(e^{\nu\, x^{2}}-1)$, and
$\nu=1$. We chose initial conditions $w(0)\approx0.999$, and $x(0)\approx0.815$.}
\end{figure}

\begin{figure}
\includegraphics[height=7cm]{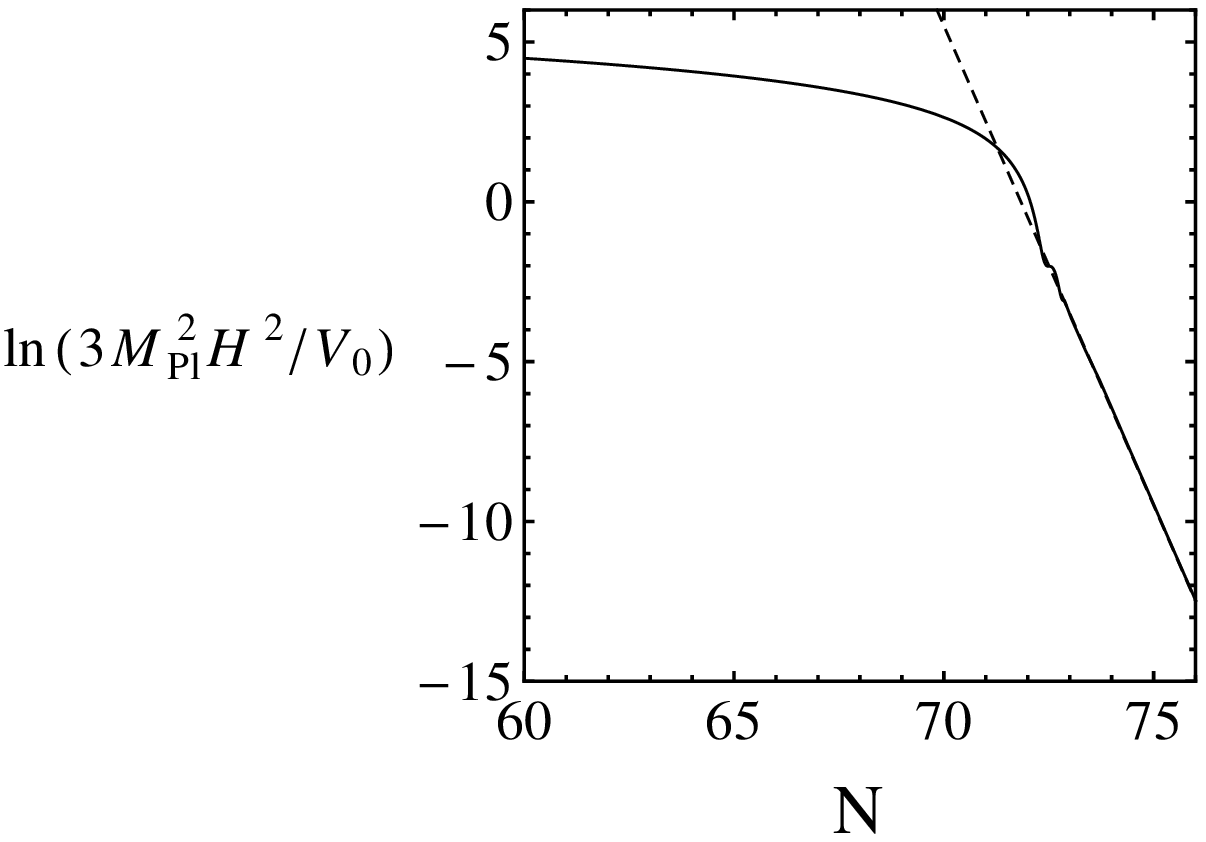} \caption{\label{fig:gs-3} The evolution of $3\Mpl^{2}H^{2}/V_{0}$ for the
potential $V=V_{0}(e^{\nu x^{2}}-1)$, and $\nu=1$ (continuous black
curve). This figure shows that after inflation ends (around $N\simeq72$),
the universe enters a matter dominated epoch, as the curve approaches
a dashed line, which represents the line $\ln(3\Mpl^{2}H^{2}/V_{0})=-3N+{\rm constant}$.
This means that after inflation $H^{2}\propto e^{-3N}\propto a^{-3}$.}
\end{figure}

It is easy to see that the effective potential has the extremum at
$x=\pm\sqrt{2/3}$ and $x=0$. Similar to the analysis for the previous
potentials, the field can drive inflation when we initially put it
in the region satisfying the condition $\epsilon\ll1$, e.g. $|x|\gtrsim\sqrt{2/3}$
and $|w|\sim1$. The condition $|\eta|\ll1$ will be satisfied when
the field is frozen nearly $x=\pm\sqrt{2/3}$. Since $x=\pm\sqrt{2/3}$
are not stable fixed points, the field can continuously evolve through
$x=\pm\sqrt{2/3}$ and then oscillates about $x=0$ eventually. This
behavior is also shown by using numerical integration methods as seen
in Figure \ref{fig:gs-2}. Because of this behavior, the speed of
propagation will be approximately equal to $c_{X}^{2}\approx1+4\nu/3$
in the slow-roll regime, whereas $c_{X}^{2}\to1$, as $x\to0$. In
Figure \ref{fig:gs-3}, we also show the behavior of the Hubble parameter
during the oscillatory regime, confirming that a matter-dominated
era takes place during this epoch.

\begin{figure}
\includegraphics[height=7cm]{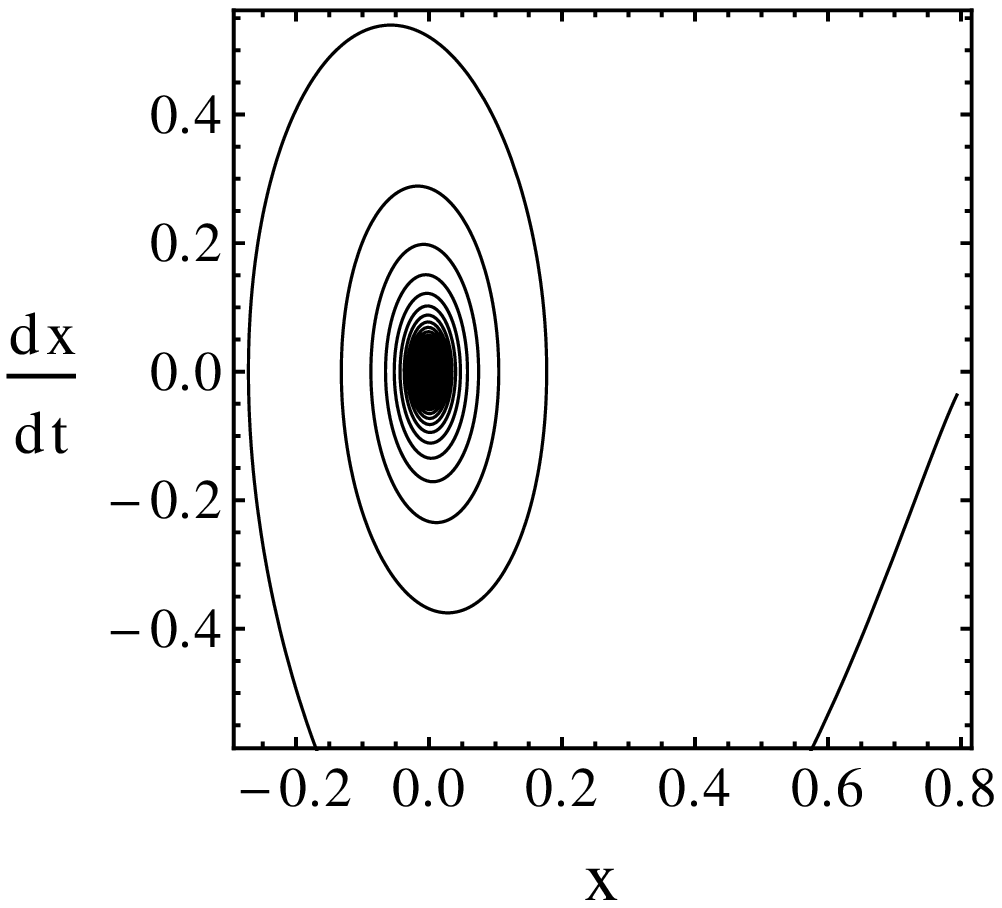}\,\,\,\,\,\,\includegraphics[height=7cm]{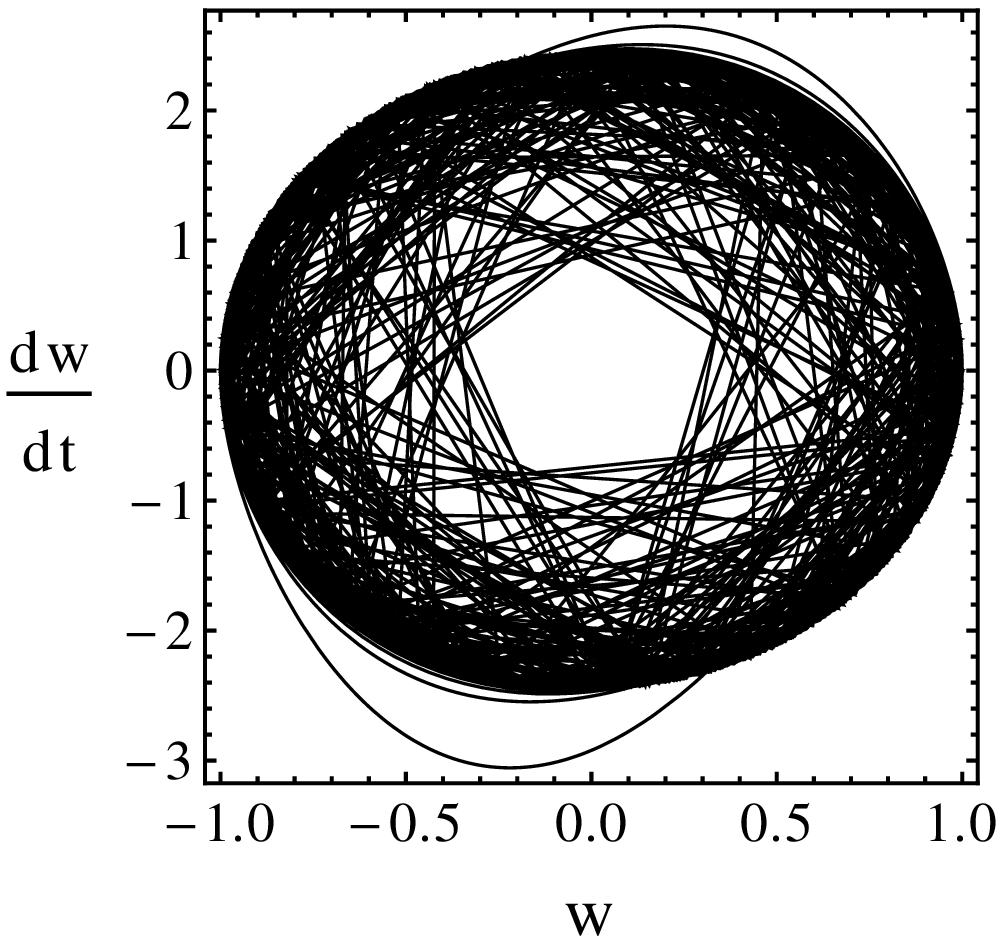}
\caption{\label{fig:gs-4}Phase space plot for the variables $dx/dt$ and $x$
(left panel), and for $dw/dt$ and $w$ (right panel), during the
oscillatory epoch.}
\end{figure}

Finally, we show the trajectory of $dx/dt$ and $x$, together with
$dw/dt$ and $w$ in Figure \ref{fig:gs-4}.


\subsection{General form of potential}

From the investigation of the previous subsections, one can see that
the viable 3-form models can be characterized by the shape of their
potential. The study of power-law potentials suggests that the viable
potential form which is free from ghosts and Laplacian instability
should have the local minimum at $x=0$. This is because when $V_{,x}$
changes sign around the minimum point (as $V$ is, by construction,
an even function of $x$, $V(x)=V(-x)$), $x$ also changes sign such
that $V_{,x}/x=12V_{,y}$ is always positive (where $y=6X^{2}$, and
$x=X/\Mpl$). In this situation, $c_{X}^{2}>0$ around $x=0$ because
$V_{,xx}>0$. The speed of propagation is still positive as long as
$x$ remains significantly far from the nearest local maximum (if
it exists) of the potential along the trajectory of motion. Hence,
if the bare potential has no local maxima between $x=\pm x_{s}$,
where $x_{s}$ is the initial value of $x$, the field can evolve
between $x=\pm x_{s}$ without giving rise to ghosts or Laplacian
instabilities. In order to avoid the stable fixed point at $x=0$,
providing the oscillation phase at this point, our investigation also
suggests that the value of the potential should be zero, $V(x=0)=0$.

We have introduced a class of potentials which are always free of
instabilities by constructions. However, this is not the only possibility.
In fact, there might be regions of the potentials which can lead to
instabilities, nonetheless those same regions are never reached by
the dynamics. This fact, can in principle, enlarge the possible inflationary
scenarios for these models, especially when we look for particular
predictions on some inflationary observables.

In other words, one can search for potentials which may give rise
to some interesting behavior of inflaton. For example, some models
of inflation can provide a possibility to generate non-Gaussianities
in CMB data. The non-Gaussianities can be characterized by a parameter
$f_{\mathrm{NL}}$ which, at least for scalar-tensor theories, can
lead to observable signatures, whenever the speed of propagation for
the field $c_{X}^{2}$ is positive but less than unity. In most of
the single-field models studied so far, the smaller $c_{X}^{2}$,
the larger $f_{\mathrm{NL}}$ \cite{gen2nd}.

Although a more detailed study is necessary to determine $f_{\mathrm{NL}}$
for 3-forms, it is interesting to see whether stable and ghost-free
3-forms can lead to a small speed of propagation $c_{X}^{2}$. For
the potential we have investigated so far, such as $V=V_{0}(x^{4}+bx^{2})$,
if we allow negative sign of the first term, the non-Gaussianities
may be generated. Let us consider, as an example, the potential $V=V_{0}\,\tanh(\nu x^{2})$
where $\nu$ is a positive constant parameter.

\begin{figure}
\includegraphics[height=7cm]{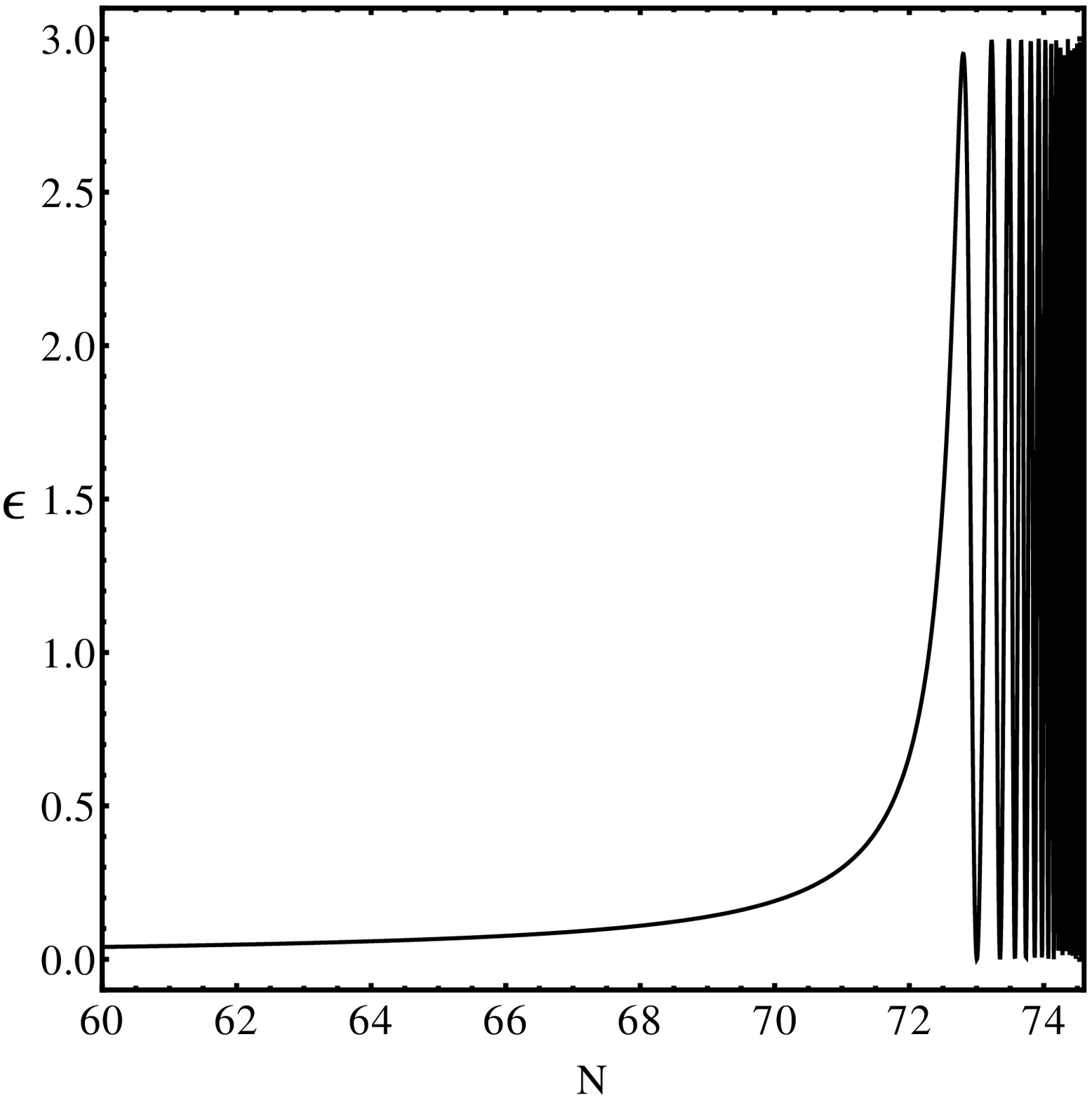}\,\,\,\,\,\,\includegraphics[height=7cm]{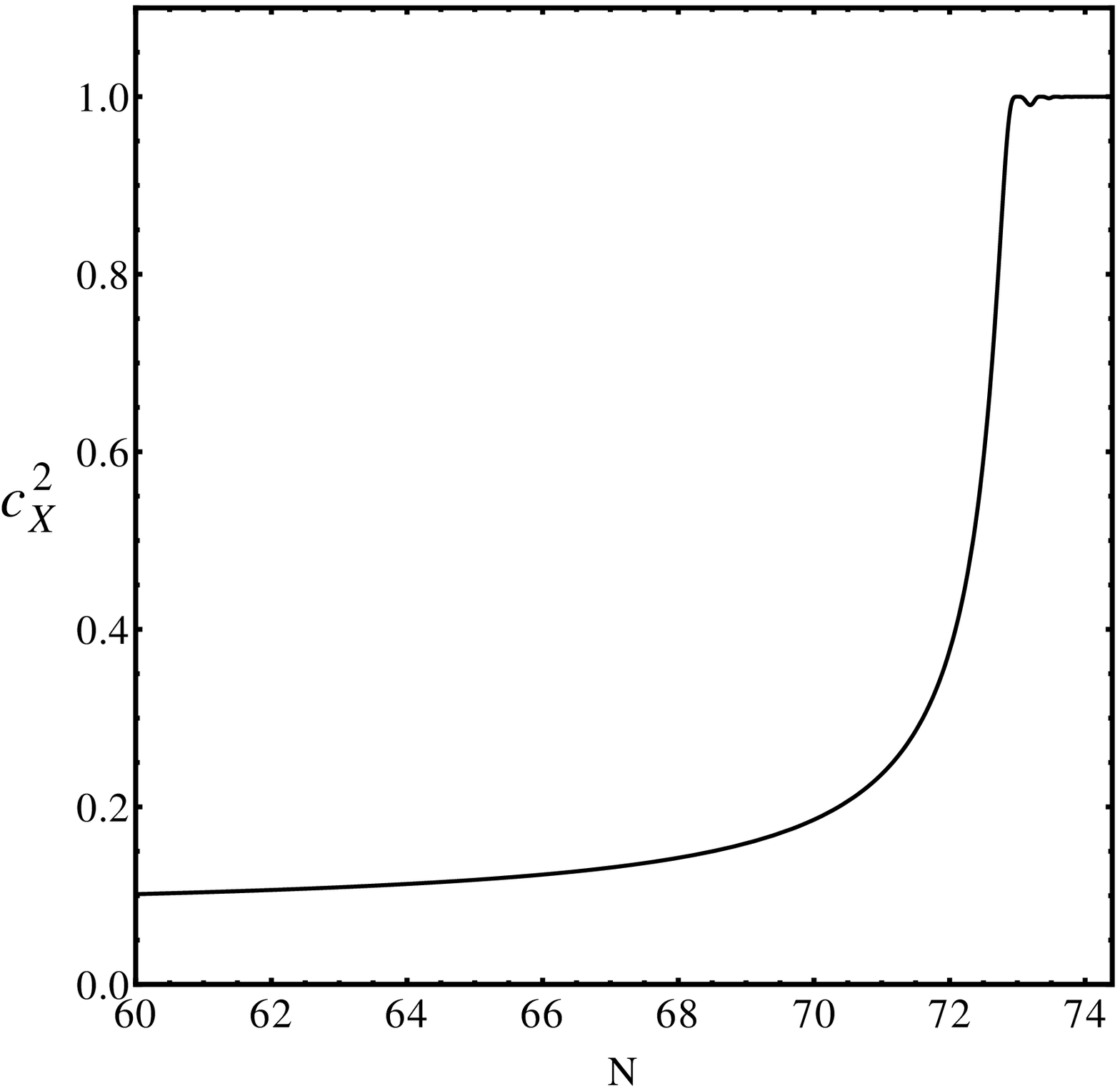}
\caption{\label{fig:tanh-1}In the left panel, this plot shows the evolution
of slow-roll parameter, $\epsilon$, for potential $V=V_{0}\,\tanh(\nu x^{2})$
with $\nu=3/4$. In the right panel, this plot shows the evolution
of speed of propagation, $c_{X}^{2}$, for potential $V=V_{0}\,\tanh(\nu x^{2})$
with $\nu=3/4$.}
\end{figure}

For the potential $V=V_{0}\,\tanh(\nu x^{2})$, the propagation speed
takes the form
\begin{equation}
c_{X}^{2}=1-4\nu\tanh(\nu x^{2})x^{2}.
\end{equation}
There is a region for the parameters in which the model is not viable
(as $c_{X}^{2}<0$). The condition for excluding this region depends
on the values of $X$ and $\nu$. If we demand the model to be viable
in the region inside $X<\Mpl$, one can set $\nu=0.52$. Then we obtain
$c_{X}^{2}(x\approx\sqrt{2/3})\approx0.534$. Smaller values for $c_{X}^{2}$
will be obtained by restricting the viable region narrower, nearly
the fixed point $x=\sqrt{2/3}$. For example, we obtain $c_{X}^{2}\sim0.076$
during inflation when we set $\nu=3/4$ as shown in Figure \ref{fig:tanh-1}.
Therefore the speed of propagation, $c_{X}^{2}$, can be small (but
positive) during inflation, however, finally, $c_{X}^{2}\sim1$ during
the oscillating phase, as expected.

There are other possible potential forms which can give the speed
of propagation less than one such as $V=V_{0}\,(x^{2}-bx^{4}+\epsilon x^{6})$
with small $\epsilon$. However, the results are not significantly
different from the form we have investigated here. We note that the
suitable form of the potential which provides small enough $c_{X}^{2}$
satisfies the condition $yV_{,yy}/V_{,y}\sim\text{constant}$ during
$\sqrt{2/3}<x<1$.

\section{General considerations and conclusions\label{sec:fine}}

We have proposed a class of potentials which are free of instabilities,
can drive inflation, and provide a final stage of matter-dominated-like
oscillatory epoch, during which reheating can occur. In order to avoid
a ghost and instabilities, these potentials should have a local minimum
at $x=0$ and have no local maximum along the trajectory of motion.
The three-form field $x$ can oscillate around this minimum if the
potential vanishes at $x=0$, i.e., the fixed point $M=(x,w)=(0,0)$
is unstable. A simple example for such a potential is
\begin{equation}
V=V_{0}\,\bigl(bx^{2}+(x^{2})^{p}\bigr)\,,\qquad{\rm with}\qquad b>0,\; p\geq1\,.\label{eq:goodp}
\end{equation}
We have introduced this form for the potential because, for simple
power-law monomials, i.e.\ $V\propto y^{p}$, with $p>1$, the second
order action for the perturbations given in Eq.~(\ref{action-s})
will vanish at $y=0$ since $Q\propto V_{,y}=0$. This corresponds,
in general, to a strong coupling limit for the theory. One can avoid
this situation by modifying the power-law potential as in Eq.~(\ref{eq:goodp}).
There is no fixed-point at $x=0$ for this form of the potential.
Therefore, the field can oscillate around $x=0$ to provide the mechanism
to end the inflation without reaching $Q=0$ at $y=0$. More in detail,
according to the previous section, the points in region $|x|>\sqrt{2/3}$
(unstable slow-roll fixed point of the dynamical equations of motion)
will be forced to move to the region $|x|<\sqrt{2/3}$, and the inflationary
period will be long enough if the field $x$ starts at $|x|=x_{s}>\sqrt{2/3}$
with $|w|\sim1$, where $w\propto x'+3x$. The bottom line is that,
in general, the 3-form field can drive long enough inflation without
the ghosts or instabilities if its potential has local minimum at
$x=0$ and has no local maximum between $x=\pm x_{s}$.

We also give another working example, the Gaussian potential, here
defined as
\begin{equation}
V=V_{0}(e^{\nu x^{2}}-1)\,,
\end{equation}
which has similar properties to the power-law case discussed above.
In fact, a long-enough slow-roll regime is followed by an oscillatory
epoch where inflation ends.

Even if avoiding ghosts ($Q>0$) and Laplacian instabilities ($c_{X}^{2}<0$)
are necessary conditions to be satisfied, they are not, however, sufficient,
in general, to have a successful period of inflation. In other words,
it is not assured that inflation ends for other classes of potentials
which are, on the other end, free from instabilities.

If the potential $V(y)$ is such that for $y\geq0$, it satisfies
the conditions $V\geq0$, $V_{,y}>0$, and $V_{,yy}\geq0$, then no
instabilities arise, as already said. However, if we also impose that
as $y\to0$, we have $V(y)\simeq c\, y$, where $c$ is a positive
constant, then for $x\approx0$ (and this point can be reached), $V_{{\rm eff}}\propto x^{2}$,
so that in general, an oscillatory epoch can take place, ending inflation.

In the last subsection in section \ref{sec:potential}, a possibility
to find non-Gaussianities from three-form model of inflation is investigated.
Our results show that some potential forms can provide the small enough
speef of propagation for the scalar modes, $c_{X}^{2}$. However,
in order to achieve small values for $c_{X}^{2}$ and to keep at the
same time a stable evolution, we had to restrict the allowed interval
for the field dynamics such that $X\lesssim\sqrt{2/3}\Mpl$.

We have investigated the stability of the perturbations for a minimally
coupled 3-form, whose action has a standard kinetic term and a generic
potential function. We have found the conditions for which the inflationary
dynamics can be stable, and gave some classes of potentials which
can provide enough inflation without generating ghosts or Laplacian
instabilities. We will leave the question to constrain the parameter
space for this potentials by using the bounds on the spectral index
and tensor-to-scalar ratio to a future research project.
\begin{acknowledgments}
We thank A.~Chatrabhuti and A.~Ungkitchanukit for their kind
hospitality at Chulalongkorn University, where part of this project
was written and terminated. We also thank S.~Tsujikawa for interesting
discussions. We especially thank referee for useful comments. K.K.~is
supported by Thailand Research Fund (TRF) through grant RSA5480009.
\end{acknowledgments}
\appendix

\section{The dual theory}

It is possible to define the 1-form or vector dual to the 3-form as
\begin{equation}
A_{\alpha\beta\gamma}=E_{\alpha\beta\gamma\delta}B^{\delta}=\sqrt{-g}\epsilon_{\alpha\beta\gamma\delta}B^{\delta}\,,\label{eq:dual1}
\end{equation}
where $E_{\alpha\beta\gamma\delta}$ is the Levi-Civita antisymmetric
tensor on curved backgrounds, which on Minkowski reduces to $\epsilon_{\alpha\beta\gamma\delta}$
(with $\epsilon_{0123}=1=-\epsilon^{0123}$). Then we also have $E^{\alpha\beta\gamma\delta}=\epsilon^{\alpha\beta\gamma\delta}/\sqrt{-g}$.
It is easy to show that $\nabla_{\mu}E_{\alpha\beta\gamma\delta}=0$.
In the following we will make use of the following relations $\epsilon^{\alpha\beta\gamma\delta}\epsilon_{\mu\beta\gamma\delta}=-6\delta_{\mu}^{\alpha}$,
and $\epsilon^{\alpha\beta\gamma\delta}\epsilon_{\mu\nu\gamma\delta}=-2(\delta_{\mu}^{\alpha}\delta_{\nu}^{\beta}-\delta_{\nu}^{\alpha}\delta_{\mu}^{\beta})$.
Therefore we obtain
\begin{equation}
A_{\alpha\beta\gamma}A^{\alpha\beta\gamma}=\epsilon_{\alpha\beta\gamma\delta}\epsilon^{\alpha\beta\gamma\mu}B_{\mu}B^{\delta}=-6B_{\mu}B^{\mu}\,.
\end{equation}

We also have
\begin{eqnarray}
-\frac{1}{48}\, F_{\alpha\beta\gamma\delta}F^{\alpha\beta\gamma\delta} & = & -\frac{1}{2}\, F_{0123}F^{0123}=-\frac{1}{2}\,(\epsilon_{1230}\nabla_{0}B^{0}-\epsilon_{0123}\nabla_{3}B^{3}+\epsilon_{3012}\nabla_{2}B^{2}-\epsilon_{2301}\nabla_{1}B^{1})\nonumber \\
 & = & {}\times(\epsilon^{1230}\nabla^{0}B_{0}-\epsilon^{0123}\nabla^{1}B_{1}+\epsilon^{3012}\nabla^{2}B_{2}-\epsilon^{2301}\nabla^{1}B_{1})=\frac{1}{2}\,(\nabla^{\mu}B_{\mu})^{2}\,,
\end{eqnarray}
so that the action is equivalent to the following one
\begin{equation}
S=\int d^{4}x\sqrt{-g}\left[\frac{\Mpl^{2}}{2}\, R+\frac{1}{2}\,(\nabla^{\mu}B_{\mu})^{2}-V(B_{\mu}^{2})\right],
\end{equation}
which shows that the 3-form action is classically equivalent to a
particular class of vector-tensor theories. Relation (\ref{eq:dual1})
can be inverted to give
\begin{equation}
B^{\mu}=\frac{1}{3!}\frac{1}{\sqrt{-g}}\,\epsilon^{\mu\alpha\beta\gamma}A_{\alpha\beta\gamma}\,,
\end{equation}
therefore, once the tensor $\bm{A}$ is known we can uniquely find
$\bm{B}$. At the level of the perturbations we find
\begin{equation}
\delta B^{\mu}=\frac{1}{3!}\frac{\epsilon^{\mu\alpha\beta\gamma}}{\sqrt{-g}}\left[\frac{A_{\alpha\beta\gamma}}{2}\, g_{\rho\sigma}\delta g^{\rho\sigma}+\delta A_{\alpha\beta\gamma}\right],
\end{equation}
which is valid on any background.

\end{document}